\documentclass{aa}  

\usepackage{graphicx}
\usepackage{txfonts}
\usepackage{float}
\usepackage{hyperref}
\begin{document}

   \title{The subsurface habitability of small, icy exomoons}


   \author{J.N.K.Y. Tjoa
          \inst{1}\thanks{Currently at the Max Planck Institute for Solar System Research, Justus-von-Liebig-Weg 3, 37077 G{\"o}ttingen, Germany; correspondence address \url{tjoa@mps.mpg.de}.}
          \and
          M. Mueller
          \inst{1,2,3}
          \and
          F.F.S. van der Tak
          \inst{1,2}
          }

   \institute{Kapteyn Astronomical Institute, University of Groningen,
             Landleven 12, 9747 AD Groningen, The Netherlands
         \and
             SRON Netherlands Institute for Space Research,
             Landleven 12, 9747 AD Groningen, The Netherlands
        \and
             Leiden Observatory, Leiden University,
             Niels Bohrweg 2, 2300 RA Leiden, The Netherlands
             }

   \date{Received xxxx; accepted xxxx}

 
  \abstract
   {Assuming our Solar System as typical, exomoons may outnumber exoplanets. If their habitability fraction is similar, they would thus constitute the largest portion of habitable real estate in the Universe. Icy moons in our Solar System, such as Europa and Enceladus, have already been shown to possess liquid water, a prerequisite for life on Earth.}
   {We intend to investigate under what thermal and orbital circumstances small, icy moons may sustain subsurface oceans and thus be ``subsurface habitable". We pay specific attention to tidal heating, which may keep a moon liquid far beyond the conservative habitable zone.}
   {We made use of a phenomenological approach to tidal heating. We computed the orbit averaged flux from both stellar and planetary (both thermal and reflected stellar) illumination. We then calculated subsurface temperatures depending on illumination and thermal conduction to the surface through the ice shell and an insulating layer of regolith. We adopted a conduction only model, ignoring volcanism and ice shell convection as an outlet for internal heat. In doing so, we determined at which depth, if any, ice melts and a subsurface ocean forms.}
   {We find an analytical expression between the moon's physical and orbital characteristics and the melting depth. Since this expression directly relates icy moon observables to the melting depth, it allows us to swiftly put an upper limit on the melting depth for any given moon. We reproduce the existence of Enceladus' subsurface ocean; we also find that the two largest moons of Uranus (Titania \& Oberon) could well sustain them. Our model predicts that Rhea does not have liquid water.}
   {Habitable exomoon environments may be found across an exoplanetary system, largely irrespective of the distance to the host star. Small, icy subsurface habitable moons may exist anywhere beyond the snow line. This may, in future observations, expand the search area for extraterrestrial habitable environments beyond the circumstellar habitable zone.}


   \maketitle
%

\section{Introduction}
\label{sec:intro}
The first exomoon candidate was recently announced \citep{TeachKip2018}, but the hunt for the first confirmed exomoon is still ongoing. Regardless of confirmation (different interpretations of the data exist: e.g., \citealt{Heller2019}), the question of exomoon habitability has now arisen. \cite{Reynolds1987} first suggested a tidally induced habitable zone might exist around gas giant planets; this tidal habitability has been extensively studied in recent years (e.g., \citealt{DobTurn2015}, \citealt{ForganDobos2016}). \cite{Scharf2006} investigated where habitable exomoon orbits might exist around giant planets and explored the possibility of maintaining temperate worlds using tidal heating; \cite{Kaltenegger2010} investigated whether biomarkers could be observed in transiting exomoon atmospheres, putting limits on detectability using Earth as a proxy. \cite{HellerBarnes2013} analyze the habitability of exomoons as constrained by their energy budgets and find that the circumstellar habitable zone for moons extends farther out than for planets. This paper intends to further investigate under what circumstances (exo)moons may sustain subsurface habitable environments, as are thought to exist on various Solar System moons (such as Europa, e.g., \citealt{Reynolds1987}; Ganymede, \citealt{Kivelson2002}; Callisto, \citealt{Khurana1998}; Enceladus, \citealt{Porco2006}; Titan, \citealt{Baland2011}; and Triton, Rhea, Titania, Oberon etc., \citealt{Hussman2006}). \\
\indent For exoplanets, the habitable zone is defined as the circumstellar region where liquid water may exist on the surface, without provoking a runaway greenhouse effect. Hence, habitability is primarily a temperature criterion and chiefly dependent on stellar illumination (see \citealt{Kaltenegger2017} for a review of planetary habitability), though many other criteria may apply \citep{Schwieterman2019}. The planet must also be massive enough to sustain an atmosphere with pressure above the triple point of water to prevent it from escaping into space. For the purposes of this paper, this condition is labeled surface habitability and conforms to the following criteria: first, the planet's or moon's surface temperature is typically between the melting and boiling points of water, that is, its surface conditions allow for the presence of liquid water; and second, the planet or moon is massive enough to maintain an (appreciable, not trace) atmosphere, but not so massive as to become a gas giant, preventing the volatile water from sublimating and escaping into space. \\
\indent A moon in the circumstellar habitable zone, if massive enough, might well be surface habitable. The major moons of our Solar System are however all airless (except Titan), beyond the surface habitable zone or both. Hence, while most moons possess water ice, they are not surface habitable. However, liquid water does not need to be on the surface to be hospitable to life. Earth's deepest oceans are active habitats: plenty of organisms thrive in underground lakes or near hydrothermal vents (see \citealt{Martin2008} for a review). Hence, liquid water below the surface may too constitute a habitable environment; in the absence of an atmosphere, a solid crust or shell must then prevent the volatile water from escaping. We label this condition subsurface habitability, and it conforms to the following criteria: first, the planet's or moon's surface temperature is typically below the sublimation point of water ice ($\sim$150 K), that is, its surface is frozen; and second, sufficient internal heat allows a global shell or pockets of liquid water to exist underneath the frozen surface. \\
\indent Following these two definitions Earth is the only object known to be surface habitable, while objects such as Enceladus, Europa or Ganymede may be subsurface habitable. Since this paper treats icy, airless moons, habitability means subsurface habitability unless otherwise specified. We will now discuss important contributions to the energy budget of icy moons, and thus key influences on subsurface habitability. \\
\indent Tidal heating introduces an important contribution to the energy budget of moons. It may render an otherwise frozen satellite habitable, or turn an otherwise habitable one into a runaway greenhouse \citep{Heller2012,HellerBarnes2013}. For surface habitability, \cite{HellerBarnes2013} find a ``habitable edge" exclusion zone around the planet below which tidal heating is so severe it could trigger a runaway greenhouse effect, leading to a Venus-like state. This edge differs from the circumstellar habitable zone in that the zone has both a lower and an upper limit; the circumplanetary edge only describes a minimum semi-major axis below which surface habitability is thought impossible, since a moon might be surface habitable at any arbitrary greater distance from its host planet if stellar illumination allows \citep{HellerBarnes2013}. It must be noted that habitable worlds (both moons and planets) may still exist beyond the habitable zone, if tidal effects or atmospheric composition so allow: see \cite{HellerArmstrong2014} for a discussion of these ``superhabitable" worlds. \\
\indent \cite{Heller2012} also finds that M-dwarfs with masses below $0.2M_{\odot}$ can not host habitable exomoons: since the habitable zone of these stars is very small as a consequence of their low luminosity the host planet would have to orbit close in, resulting in a very small Hill sphere and thus very small moon semi-major axes. The star's tidal influence might continue to affect the habitable zone up to $0.5M_{\odot}$; subsequent eccentricity forcing by the star would see the exomoon tidally roasted. \\
\indent Tidal (or other internal) heating may also contribute to habitability by enabling geothermal activity. Hydrothermal vents have been suggested as catalysts of abiogenesis on Earth; the first life forms on Earth could have emerged here \citep{Martin2008,Dodd2017}. The geothermal activity (at least on Earth) required to sustain these environments is driven by the dissipation of internal energy; tidal heating may induce similar situations in the subsurface oceans on exomoons, providing not only a habitat but also the means by which to catalyze possible inhabitants \citep{Hsu2015}. \\
\indent Tidal heating is key to exomoon habitability since it presents a major deviation from planetary habitability as we see in our own solar system. Planetary systems, except for compact M-dwarf systems (such as TRAPPIST-1; see, e.g., \citealt{Dobos2019} and \citealt{HayMatsu2019}), may be too large to allow for significant tidal effects; the magnitude of tidal heating scales very strongly with the distance between the bodies ($\dot{E}_{tidal} \propto a^{-7.5}$, with $a$ the semi-major axis of the secondary body) and thus can not take place over large distances. Conversely, satellite systems are relatively small compared to the sizes of their constituent bodies and thus are more frequently tidally active. Examples of tidally heated moons in the Solar System include Io and Europa (by Jupiter; \citealt{Yoder1979}) and Enceladus (by Saturn; \citealt{Porco2006}). The first is a desiccated and highly volcanically active, being situated well inside Jupiter's habitable edge; conversely, Enceladus is thought to sustain a global subsurface ocean thanks to tidal heating. \\
\indent While illumination mechanisms deposit their heat on the surface, tidal heating deposits heat at depth; it can thus directly influence and thus maintain liquid possible subsurface oceans. Additionally, thermal conduction of internal heat through a kilometers thick ice shell is inefficient which means that the internally deposited tidal heat can be retained over long timescales. Finally, it provides (if the moon's eccentricity can be maintained over long timescales) a constant source of heating rather than variable exogenic flux. It must however be noted that tidal heating may also give rise to tectonics and cryovolcanism on icy moons by partially melting their interiors and inducing ocean flows \citep{Spencer2009}; these processes (convective and advective heat loss) thus provide other outlets of internal heat next to conductive heat loss through the crust. \\
\indent Since tidal heat comes at the expense of orbital energy, the tidally heated object suffers from orbital damping, circularizing its orbit and slowing its rotation rate to a tidal lock over time; however, a nonzero eccentricity is required for tidal heating to be effective \citep{Jackson2008}. Hence, a mechanism is required to maintain the eccentricity. Mean motion resonances serve this purpose: the periodic gravitational tugs experienced by the inferior partner in a small integer resonance (1:2, 1:3, 2:3 etc.) boost its eccentricity sufficiently to counteract complete orbital damping. Although resonances are the primary example of such a mechanism in the solar system (Dione-Enceladus in 1:2, Ganymede-Europa-Io in 1:2:4), superior planets or moons outside resonance are also capable of providing the necessary perturbations to counteract orbital damping, as is the nearby star in compact systems \citep{vLaerhoven2014}. \\
\indent In addition to tidal heat, moons' energy budgets also deviate from planets' in that they receive heat from, or via, the planet. The planet radiates thermally and reflects stellar emission onto the moon. \cite{HellerBarnes2013} investigate the effects of planetary illumination and find that, at a constant stellar distance, as reflected heat goes up, thermal heat must go down (since the albedo goes up). In either case, planetary illumination is often small in comparison to stellar or tidal contributions. \\
\indent \cite{ForganYotov2014} and \cite{ForganDobos2016} studied the effect of frequent stellar eclipses on the ice-albedo feedback mechanism (where the high albedo of surface ices drives an additional temperature decline) and find that, if the orbits of moon and planet are close to coplanar, this mechanism may drive the moon into a snowball state that it finds hard to escape. However, if one wishes to create a subsurface habitable environment, snowball states are acceptable as long as internal heat is sufficient to sustain an ocean layer. \\
\indent By virtue of its composition (mostly water ice) Enceladus serves as a template for the type of small, tidally heated icy moon we are interested in. It is too small to sustain a fully molten interior\footnote{In this paper, a ``molten interior" implies a mantle and core consisting of molten rock, such as silicates; a subsurface ocean is not what is meant by molten interior.}, but its high heat flux (particularly through the active south polar terrain; see \citealt{Porco2006}) implies that this interior must be substantially deformable. Were it not for tidal heating, given its distance from the Sun, Enceladus would be frozen solid; however, Enceladus is now thought to have a porous, spongiform (and thus deformable) core, allowing the water of its presumed subsurface ocean to efficiently transport the internally dissipated tidal heat upward, thus maintaining its liquidity \citep{Nimmo2018}. Evidence supporting the existence of Enceladus' subsurface ocean include its geological activity: as a consequence of its tidal heating Enceladus also displays vigorous cryovolcanism on its active south pole, feeding Saturn's E-ring with plumes of both icy particles and more complex molecules \citep{Porco2006}. In addition, the recent discovery of molecular hydrogen in these plumes suggests that hydrothermal processes are occurring in Enceladus' ocean to replenish this volatile and hints at the presence of hydrothermal vents \citep{Waite2017}. It is our intention to find under what variety of circumstances small, icy moons such as Enceladus may still sustain subsurface oceans. \\ \\
\indent This paper studies the energy budget and the influence of different heat sources on (exo)moon subsurface habitability. We determine under what circumstances exomoons can host subsurface oceans. We hypothesize that the circumstellar subsurface habitable zone for exomoons, similar to its surface counterpart, extends farther out than for planets because of additional heating by the planet and tidal effects. In addition, we hope to find a circumplanetary subsurface habitable edge similar to the one described by \cite{HellerBarnes2013} for surface habitability. \\
\indent Section \ref{sec:physb} discusses the physical background to tidal heating and conductive and radiative cooling; Section \ref{sec:appr} presents an overview of our melting depth model and its dependencies. We then apply our model to four fiducial moons (Enceladus, Rhea, Titania, Oberon) to demonstrate its use, showcasing our results in Section \ref{sec:results}, and discuss the implications our model has for icy moon habitability in Sections \ref{sec:disc} and \ref{sec:conclusions}.

\section{Physical background}
\label{sec:physb}
\subsection{Approaches to tidal heating}
\label{ssec:physb/aptoth}
Fixed Q and viscoelastic models are the most commonly used approaches to tidal heating. Though the following expression does not fully capture the intricacies of tidal heat dissipation, in both cases the total heat dissipated in the moon's interior can be estimated by:

\begin{align}
    \Dot{E}_{tidal} = \dfrac{21G^{1.5}}{2} \cdot \Phi \cdot \dfrac{M_p^{2.5} R_s^5 e_s^2}{a_s^{7.5}},
    \label{eq:tidalheating}
\end{align}

as stated in for instance \cite{Henning2009}, and wherein $M_p$ is the mass of the primary (or host/parent) and $R_s$, $e_s$, and $a_s$ are the radius, orbital eccentricity, and semi-major axis of the secondary (or satellite) body. The definition of $\Phi$, which we dub the tidal efficiency factor, varies per approach. \\
\indent Fixed Q models are phenomenological and do not describe much underlying physics. They assume that the body is uniform in composition, lumping the object's tidal response into two terms: $\Phi=Q/k_2$, where the tidal quality factor $Q$ is an inverse damping term describing the lag in the uniform body's spring response and the second order Love number $k_2$ describes the body's deformation response to stress (see \citealt{Henning2009} for a review). If the body is not entirely uniform but has a uniform interior below some depth, the quotient $Q/k_2$ is multiplied by a factor $f_{V,tidal}$, which is the volume fraction of the moon taken up by the uniform interior (i.e., everything but the crust and ocean; generally, $f_{V,tidal} \sim 0.8$). \\
\indent If a body is sufficiently nonhomogeneous (i.e., has a molten interior or differentiation, such as found in high-mass moons such as Europa), fixed Q becomes less valid and viscoelastic models must be invoked (see \citealt{RenaudHenning2018} for a recent review). These are more complex and describe more underlying physics, accounting for motion in the mantle; however, since we focus on small, icy moons that lack fully molten interiors, a simpler and phenomenological approach is sufficient. Since the constituent terms of fixed Q tidal heating are as phenomenological as $\Phi$ itself, in this work $\Phi$ is varied directly. \\
\indent Both approaches are fairly simplistic; more advanced approaches to tidal theory exist (e.g., \citealt{FerrazMello2008,Leconte2010,BoueEfroimsky2019}). However, we choose the more simplistic approach set forth in Equation \ref{eq:tidalheating}, since our goal is not to investigate in detail the tidal behavior of exomoons; we aim to gain an impression of where habitable exomoons may exist irrespective of their exact tidal behavior. Equation \ref{eq:tidalheating} gives us sufficient information in this regard, namely the total amount of energy dissipated in the body's interior as a function of its observable characteristics, and with all internal and structural properties lumped into a single factor $\Phi$. It is possible to mathematically untangle this factor into its constituent terms in any number of more detailed ways; this is left for future research. For the purposes of this work, $\Phi$ suffices. \\
\indent It must be noted that, since our approach is phenomenological, we may be assuming physically unrealistic scenarios. We do not model the direct relation between tidal heating and internal structure, nor do we model the relation between $Q$ and $k_2$ and the moon's orbit \citep{Henning2009}. We sample the parameter space for $\Phi$ without much regard as to what exactly the structural implications are, which are not the focus of this work.

\subsection{Heat conduction}
\label{ssec:physb/heattrans}
Heat is supplied to the moon via endogenic (internal) and exogenic (external) pathways. Endogenic sources include tidal heating, radiogenic heating (by decay of radioactive elements in the body), and residual accretion heat (heat trapped during body formation); exogenic sources include stellar illumination, planetary reflected illumination, and planetary thermal illumination. \\
\indent Endogenic heat is carried to the surface through conduction, convection (tectonics), and advection (volcanism). For conduction, the flux $F$ through a layer of thickness $D$ of constant thermal conductivity $k$ given a temperature difference $\Delta T$ is

\begin{align}
    F = k\dfrac{\Delta T}{D},
\end{align}

which can be rewritten for the temperature difference:

\begin{align}
    \Delta T = \dfrac{D\cdot F}{k}.
    \label{eq:tempgrad}
\end{align}

If we assume all heat loss to space at the surface to take place via radiation, then to maintain thermal equilibrium between the surface and the surrounding radiation field we have:

\begin{align}
    F_{endo} + F_{exo} = F_{out},
\end{align}

wherein $F_{endo}$ is the heat flux supplied by endogenic processes to the surface, $F_{exo}$ is the exogenic radiative heat flux absorbed by the given unit of surface area of given albedo, and $F_{out}$ is the outgoing radiated heat flux. It then follows that:

\begin{align}
    T_{surf} = \sqrt[4]{\dfrac{\bar{F}_{exo} + \bar{F}_{endo}}{\epsilon_s \sigma_B}},
    \label{eq:surftemp}
\end{align}

wherein $\epsilon$ is the emissivity of the given surface and $T_{surf}$ is that surface's temperature.

\section{Approach}
\label{sec:appr}

\subsection{Exogenic heating}
\label{ssec:appr/exomod}
Our exogenic heating model is based on \cite{HellerBarnes2013}; their section 3.1 details how orbit-averaged incident fluxes can be computed for both star and planet. The total orbit-averaged exogenic flux equals:

\begin{align}
    \bar{F}_{exo} = \dfrac{L_*\left(1 - \alpha_s\right)}{16\pi a_p^2 \sqrt{1-e_p^2}} \left[\bar{f}_{*,vis} + \dfrac{\pi R_p^2 \alpha_p}{2a_s^2} + 
    \dfrac{R_p^2\left(1 - \alpha_p\right)}{2a_s^2} x_{exc} \right],
    \label{eq:avexoflux}
\end{align}

wherein $L_*$ is the luminosity of the star, $\alpha_s$ the moon's albedo, $a_p$ the planet's semi-major axis, $e_p$ the planet's eccentricity, $R_p$ the planet's radius, $\alpha_p$ the planet's albedo, and $a_s$ the moon's semi-major axis. $x_{exc}$ is a flux excess factor which we use to approximate the way in which gas giants exceed their supposed equilibrium temperature through residual primordial heat (radiogenic, formation or contraction heat, ranging from 1 for Uranus to $\sim$2.5 for Saturn); $\bar{f}_{*,vis}$ is an orbit-averaged visibility fraction of the star from the moon because of eclipses by the planet and is given by:

\begin{align}
    \bar{f}_{*,vis} = 1 - \dfrac{R_p}{2\pi a_s} |\cos{i_s}|
    \label{eq:avvisfrac},
\end{align}

wherein $i_s$ is the moon's orbit's inclination relative to the planet's orbital plane, so the moon's inclination relative to the planet's equator plus the planet's own axial tilt. This term describes how much of its orbit the moon spends in the planet's shadow; higher inclination would see the moon lifted from the planet's shadow more often. Similar to Heller \& Barnes' Equation 22, the first term describes direct starlight, the second reflected starlight from the planet, and the third planetary thermal emission. \\
\indent Equation \ref{eq:avexoflux} differs from \cite{HellerBarnes2013} only in the last (planetary thermal) term: our expression has a factor 2 in the denominator, as opposed to their 4. This results from different assumptions about planetary surface temperature: \cite{HellerBarnes2013} assume a 100 K difference between the day and night sides of the host planet because of a possible tidal lock, whereas we assume (based upon the observed rotational periods of Solar System gas giants) that the host rotates rapidly. As such, it equally distributes energy across its surface, resulting in the same temperature on the day and night sides. Note however that this factor 2 difference makes negligible difference in the final results: planetary thermal emission is consistently among the weakest sources of heating.

\subsection{Endogenic heating}
\label{ssec:appr/endomod}
We assume all endogenic contributions are dissipated in the solid core and mantle. We adopt a phenomenological approach for tidal heating: we lump the body's tidal response into one term $\Phi$, the tidal efficiency factor, and vary that within reasonable limits. Current observations give values of $\Phi$ for Enceladus ($0.0026\leq \Phi\leq 0.0127$) and Io ($\Phi\approx 0.015$; both values cited from \citealt{Nimmo2018}). We then use Equation \ref{eq:tidalheating} to compute the generated tidal heat. We also use the following equation to compute the nontidal internal heat generated by radiogenic and residual heating processes:

\begin{align}
    \Dot{E}_{ac,rg,s} = \Dot{E}_{ac,rg,\oplus} \cdot \dfrac{M_s}{M_\oplus},
    \label{eq:radaccheat}
\end{align}

wherein $\Dot{E}_{ac,rg,\oplus}$ is the total internally generated heat of Earth (about 20 TW, as found by \citealt{JelliJackson2015}), and $M_s$ and $M_\oplus$ are the masses of the moon and Earth respectively. Therefore, we scale Earth's internal heat production to the moon. This treatment is in agreement with the value for Enceladus' background endogenic heat found by \cite{Czechowski2004}: $3.25\cdot 10^{-12}$ W kg$^{-1}$ versus $3.33\cdot 10^{-12}$ W kg$^{-1}$ via Equation \ref{eq:radaccheat}. Since we scale relative to Earth, this expression assumes the target body is of the same age as Earth; if the target body is of a significantly different age or has different isotopic ratios than Earth, this treatment either over- or underestimates the residual heat. If the target body is older than Earth, the total residual heat should be lower than Equation \ref{eq:radaccheat} indicates, while if the body is younger, residual heat should be higher. Similarly, if the body has a higher radionuclide content than Earth it should generate more endogenic heat. Comparing the values in Table 2 in \cite{NeumannKruse2019} and Table 3 in \cite{McDonough2019} shows that while Earth may be depleted in potassium compared to (what is modeled for the initial state of) Enceladus, it makes up for that in terms of uranium and thorium. In the end Equation \ref{eq:radaccheat} still yields a heat production in agreement with \cite{Czechowski2004}, so exactly what isotope produces this heat is not relevant to the results of this work. Important to note is that \cite{NeumannKruse2019} assume Enceladus is primordial, which may not be the case (see Section \ref{ssec:disc/impforfidms} and \citealt{Cuk2016}) and may influence their results for Enceladus' isotopic abundances today. We do not know whether Enceladus is representative for all icy moons in its modeled isotope abundances; regardless, until more data on this topic becomes available, we use Equation \ref{eq:radaccheat} as a baseline.

\subsection{Melting depth model}
\label{ssec:appr/meldep}

\begin{figure}
    \includegraphics[width=\columnwidth]{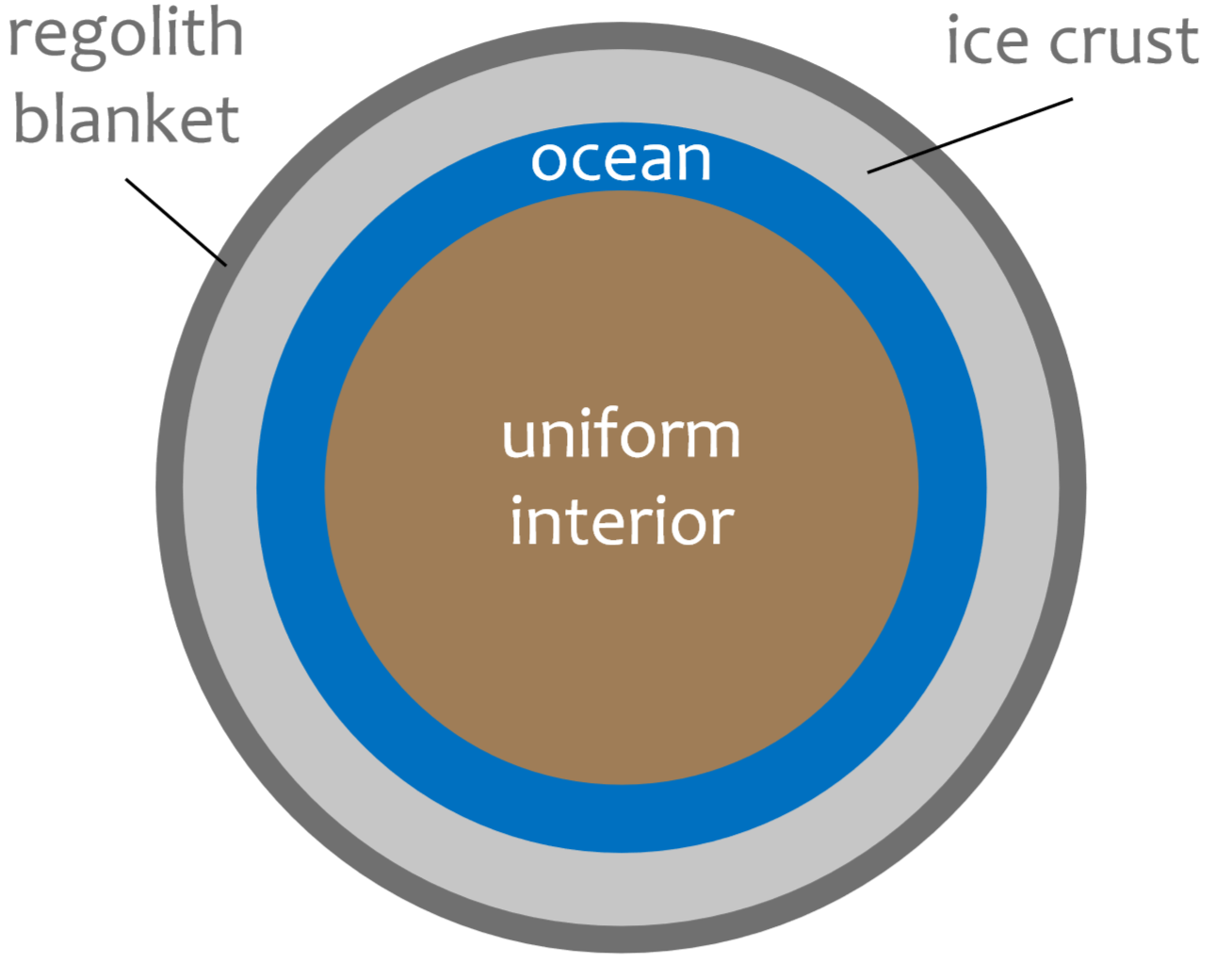}
    \caption{Schematic view of the internal structure we assume for our small, icy moons. The uniform interior is where tidal and other endogenic heat is dissipated and generated; dissipation in the ocean, ice crust, and regolith is ignored in our model. Endogenic heat is then conducted toward the surface through the ocean, ice crust, and regolith. The ocean is assumed a perfect thermal conductor, whereas the ice crust and regolith cover have finite conductivity; the regolith is highly insulating and can be thought of as a blanket. Exogenic heat is incident on the regolith blanket, with a fraction being absorbed and the rest reflected into space. \label{fig:intstruc}}
\end{figure}

Our melting depth model is fully conductive: we do not consider the heat lost through advective processes (such as volcanism or cryovolcanism). We assume the crust has two layers: the thicker, solid ice shelf below and a finely grained regolith layer of very low thermal conductivity on top (we choose 0.001 W m$^{-1}$ K$^{-1}$ based on \citealt{YuFa2016}, who determined Lunar regolith properties). Our algorithm first computes the surface temperature (Equation \ref{eq:surftemp}) as induced by some endogenic plus exogenic flux, then the temperature below the insulating regolith layer $D_{rego}$ and then the depth at which the temperature reaches the melting point of water (liquidus temperature) for some given mass fraction of ammonia (NH$_3$). \\
\indent We consider the fraction of NH$_3$ specifically since it significantly influences both the functional shape of the conductivity and the liquidus temperature. NH$_3$ is thought to be abundant in the outer Solar System, as are other pollutants that have a similar effect, notably sea salt (NaCl; \citealt{Hammond2018}). However, NH$_3$ stands out because it lowers the liquidus temperature by as much as 50 K as opposed to a maximum of $\sim$20 K for NaCl \citep{KnauthBurt2002}; \cite{Leliwa2002} find that the liquidus temperature reaches a minimum at 176.2 K for 32.6\% NH$_3$. We therefore do not consider other pollutants since NH$_3$ provides an upper limit to their possible effect. It must be noted that NH$_3$ is toxic to most terrestrial life \citep{Ip2001}, certainly at the levels potentially reached in subsurface oceans \citep{Hammond2018}; however, it is not inconceivable that local lifeforms could have evolved to metabolize NH$_3$. Further discussion of the biochemistry of possible ocean inhabitants is beyond our scope. \\
\indent In this fashion we derived an analytic expression for the melting depth (derivation shown in Appendix \ref{app:derivmeltdep}):

\begin{align}
    D_l = R_s - \left[4\pi\dot{E}_{endo}^{-1}\left(K(T_l) - K(T_{rego})\right) + \left(R_s - D_{rego}\right)^{-1}\right]^{-1},
    \label{eq:meltdepthan}
\end{align}

wherein $D_l$ is the melting depth, $R_s$ the moon's radius, $\dot{E}_{endo}$ the total generated endogenic heat, $D_{rego}$ the regolith thickness, $T_l$ the liquidus temperature, $T_{rego}$ the temperature underneath the regolith, and $K(T)$ the integrated conductivity. If we adopt the conductivity of ice $k(T)$ as found by \cite{AndersInaba2005}, namely

\begin{align}
    k(T) = 632T^{-1} + 0.38 - 0.0197T,
    \label{eq:icecond}
\end{align}

wherein $k(T)$ is the thermal conductivity of ice in W m$^{-1}$ K$^{-1}$ and $T$ the temperature in K, then $K(T)$ becomes:

\begin{align}
    K(T) = 632\ln{T} + 0.38T - 0.00985T^2.
    \label{eq:intcond}
\end{align}

For some input fiducial model, Equation \ref{eq:meltdepthan} can thus directly determine the melting depth and as such explore the parameter space within which any exomoon may exist. This is an important result, since Equation \ref{eq:meltdepthan} holds for different expressions for $k(T)$ and associated $K(T)$. Hence, if the shape of $k(T)$ is in future experiments more precisely determined, Equation \ref{eq:meltdepthan} can be easily updated. While it lacks the accuracy of more detailed, moon-specific approaches, Equation \ref{eq:meltdepthan} is highly versatile and allows us to swiftly put an upper limit on a moon's melting depth directly from icy moon observables. \\
\indent Our default value for the liquidus temperature $T_l$ is 273.15 K. $T_l$ is also a function of pressure; at lower pressure the liquidus temperature decreases until it reaches water's triple point at $\sim$250 K. However, icy moons may also possess sufficient NH$_3$ to influence both the functional shape of the conductivity and the liquidus temperature. It must be noted that, according to \cite{Hammond2018}, the NH$_3$ preferentially ends up at the bottom of the ice shelf (where the ice is partially molten) and in the subsurface ocean, so we do not take possible changes to $k(T)$ of the ice shelf into account. We do adjust the liquidus temperature dependent on the mass fraction of NH$_3$ $f_{m,NH_3}$; we used the following linear interpolation between 0 and 32.6\% NH$_3$:

\begin{align}
    T_l(f_{m,NH_3}) = \dfrac{176.2 - 273.15}{0.326}\cdot f_{m,NH_3} + 273.15.
\end{align}

In reality, \cite{Dodson2009} find the NH$_3$ fraction in the protosolar nebula to be at most around 15\%, so we adopt that as the maximum used value. We set a maximum melting depth of 30\% of the moon's radius based on estimates by \cite{Hussman2006}. We also assume a minimum melting depth of 100 m underneath the regolith layer, since the regolith has to rest on something. If our expression yields a melting depth smaller than 100 m we assume the moon to be desiccated; without a massive ice shell the volatile water quickly escapes into space. If the melting depth is larger than $0.3R_s$ we assume the moon is frozen solid. Both limits are arbitrary and could be set differently; an exceedingly large melting depth does however limit how much tidal heat may still be dissipated in the uniform interior, which would then become increasingly small (tidal heat could still be deposited in the ocean or crust, but such processes are beyond our model).

\subsubsection{Model dependencies}
\label{sssec:appr/meldep/moddep}
Our model contains the following seventeen parameters:

\begin{itemize}
    \item The stellar luminosity $L_*$: the brighter the star, the higher the moon's surface temperature.
    \item The planet's mass $M_p$: the more massive the planet, the stronger the tidal forces.
    \item The planet's semi-major axis $a_p$: the farther from the star, the lower the moon's surface temperature.
    \item The planet's eccentricity $e_p$: the more eccentric the planet's orbit, the more variable the moon's surface temperature.
    \item The planet's Bond albedo $\alpha_p$: the darker the planet, the less reflection hits the moon.
    \item The planet's emissivity $\epsilon_p$: the lower the emissivity, the less planetary emission hits the moon.
    \item The planet's flux excess $x_{exc}$: the lower the flux excess, the lower the planet's thermal flux.
    \item The moon's mass $M_s$: the more massive the moon, the stronger the tidal forces.
    \item The moon's density $\rho_s$: the denser the moon, the weaker the tidal forces.
    \item The moon's semi-major axis $a_s$ (abbreviated in plots as SMA): the farther from the planet, the weaker the tidal forces.
    \item The moon's eccentricity $e_s$: the higher the eccentricity, the stronger the tidal forces.
    \item The moon's inclination $i_s$: the higher the inclination, the fewer planetary eclipses.
    \item The moon's Bond albedo $\alpha_s$: the darker the moon, the higher the surface temperature.
    \item The moon's emissivity $\epsilon_s$: the lower the emissivity, the longer the moon takes to cool.
    \item The moon's regolith thickness $D_{rego}$: the thicker the regolith blanket, the warmer the moon's interior.
    \item The moon's tidal efficiency factor $\Phi_s$: the higher the tidal efficiency, the stronger the tidal heating.
    \item The moon's NH$_3$ mass fraction $f_{m,NH_3}$: the higher the NH$_3$ content, the lower the melting point of ice.
\end{itemize}

Exogenic heating processes also depend on the radius of the planet, but gas giant density and radius have been found by \cite{ChenKip2017} to scale with mass; therefore, we used planet mass as a proxy for planet radius. This relation is given by

\begin{align}
    \log_{10}\left(\dfrac{R}{R_\oplus}\right) = \begin{cases} 
    0.3756 + 0.589\cdot\log_{10}\left(\dfrac{M}{M_\oplus}\right) \\
    \text{for  } 2.04 M_\oplus \leq M \leq 0.414 M_{Jup} \\ \\
    83.6663 - 0.044\cdot\log_{10}\left(\dfrac{M}{M_\oplus}\right) \\
    \text{for  } 0.414 M_{Jup} \leq M \leq 0.0800 M_\odot.
    \end{cases}
    \label{eq:ChenKipmasstrend}
\end{align}

Of these seventeen parameters, twelve strongly influence the melting depth, with the moon's orbital inclination and emissivity and the planet's albedo, emissivity, and flux excess being of minor influence. The planet's eccentricity only begins to strongly influence the melting depth for values greater than 0.1, which renders acquisition of an extensive satellite system unlikely a priori. In addition, such high eccentricities do not occur for icy moon hosts in the Solar System. Icy moon densities vary only between about 1 and 2 g cm$^{-3}$, a range which does not introduce large changes in our results. Icy moon albedo, while varying widely from moon to moon, influences how far beyond the snow line it can maintain its icy shell and not so much its melting depth. The tidal efficiency factor and NH$_3$ mass fraction are not well constrained; however, because of its importance, the tidal efficiency factor is varied in our model. We do stick to fiducial values for the NH$_3$ mass fraction, reducing our model to the following eight key variables: $L_*$, $M_p$, $a_p$, $M_s$, $a_s$, $e_s$, $D_{rego}$, and $\Phi$. We then varied these parameters against one another and computed the corresponding melting depths in terms of the moon's radius. Their ranges, plus corresponding scientific questions and justification for their limits, are:

\begin{itemize}
    \item $10^{-2} L_\odot \leq L_* \leq 10^{2} L_\odot$: can subsurface oceans exist on icy exomoons around different stellar types? This range stretches from the faintest M dwarfs to B stars. This covers essentially the full main sequence except O stars, which are so luminous and live so briefly that planet formation, let alone the development of a habitable environment if not life, is unlikely.
    \item $10^{25}$ kg $\leq M_p \leq 10^{28}$ kg: can subsurface oceans exist on icy exomoons around more massive gas giants, or worlds smaller than Neptune? This range stretches from roughly 2 Earth to 10 Jupiter masses. 2 Earth masses is the ``Neptunian world" cutoff described by \cite{ChenKip2017}, above which planets acquire extensive gaseous envelopes; 10 Jupiter masses is near the canonical cutoff point for brown dwarfs as described by \cite{Spiegel2011}.
    \item 1 AU $\leq a_p \leq 10^{2}$ AU: how does the melting depth depend on stellar proximity and thus stellar illumination? This range stretches from the Solar habitable zone to roughly twice the distance of the Kuiper Belt. Inward the habitable zone, habitability is impossible a priori; beyond the upper limit, stellar illumination plays little role, hence greater distance from the star makes very little difference. We did not select the Solar snow line as the inner bound since our stellar luminosity goes down to M dwarfs, whose snow line is tighter.
    \item $10^{19}$ kg $\leq M_s \leq 10^{22}$ kg: can subsurface oceans exist on more or less massive exomoons? This range stretches from 0.1 Enceladus to roughly 3 Titania masses. 0.1 Enceladus masses is close to the mass of Mimas, the smallest gravitationally rounded body in the Solar System; above several Titania masses, we enter the regime of the Galilean moons, which possess the molten interiors our model does not apply to.
    \item $10^{8}$ m $\leq a_s \leq 10^{9}$ m: is there a circumplanetary, subsurface habitable edge or zone and if so, where is it located? This range stretches from roughly 2 Saturn radii to about the semi-major axis of Titan. The lower limit is close to the fluid Roche limit; beyond the upper limit, tidal heating no longer plays an appreciable role, hence greater distance from the planet makes little difference.
    \item $10^{-4} \leq e_s \leq 10^{-1}$: how dependent is the melting depth on eccentricity and are eccentricities critical to maintaining a subsurface ocean? This range stretches from essentially circular to half the eccentricity of Mercury ($\sim$0.2). Most regular moon orbits have eccentricities around $10^{-2}$; however, in hypothetical systems with more eccentricity pumping (by either mean motion resonance with neighboring moons or, in M dwarf systems, the nearby star or planets) this might still go up. The $10^{-1}$ upper limit is arbitrary and bears no physical meaning; higher eccentricities are possible but would likely not provide any further insight.
    \item 1 m $\leq D_{rego} \leq 10^{2}$ m: how does the melting depth depend on surface insulation and can subsurface oceans exist if the solid, icy crust is directly exposed? This range stretches from lunar levels to 100 times that. Above 100 m we assume gravity compresses the regolith to a solid crust; we consider less than 1 m of finely grained material unlikely.
    \item $10^{-4} \leq \Phi \leq 10^{-1}$: how strongly is the melting depth dependent on tidal efficiency? This range varies from essentially zero efficiency ($\Phi$ for Enceladus is 50 times our lower limit) to high efficiency (our maximum $\Phi$ is ten times what literature gives for Io). Since this parameter is poorly constrained, these limits are fairly arbitrary.
\end{itemize}

\begin{table*}
	\begin{center}
		\caption{Planets (left) and moons (right) used as fiducial models. We note that 1) emissivities are average estimates; 2) the values for $x_{p,exc}$ are estimates; and 3) the inclinations listed here are equal to the inclinations of the moons relative to their host planets' equators plus the hosts' axial tilts.}
		\begin{tabular}{ l  l  l }
		\hline\hline
		 & Saturn & Uranus \\
		\hline
		$M_p$ ($M_\oplus$) & 95.159 & 14.536 \\
		$R_p$ (10$^6$ m) & 58.232 & 25.362 \\
		$\alpha_p$ & 0.342 & 0.3 \\
		$x_{p,exc}$ & 2.5 & 1.1 \\
		$a_p$ (AU) & 9.5826 & 19.2184 \\
		$e_p$ & 0.0565 & 0.0464 \\
		$i_p$ (deg) & 2.485 & 0.773 \\
		\hline
		\end{tabular}
        \qquad
		\begin{tabular}{ l  l  l  l  l}
		\hline\hline
		 & Enceladus & Rhea & Titania & Oberon \\
		\hline
		host planet & Saturn & Saturn & Uranus & Uranus \\
		$M_s$ (10$^{20}$ kg) & 1.08 & 23.06 & 35.27 & 30.14 \\
		$\rho_s$ (g cm$^{-3}$) & 1.61 & 1.24 & 1.72 & 1.63 \\
		$\alpha_s$ & 0.81 & 0.70 & 0.17 & 0.14 \\
		$\epsilon_s$ & 0.95 & 0.90 & 0.60 & 0.60 \\
		$\Phi_s$ & 0.005 & 0.009 & 0.010 & 0.010 \\
		$D_{rego}$ (m) & 20 & 50 & 50 & 50 \\
		$f_{m,NH_3}$ & 0.00 & 0.10 & 0.15 & 0.15 \\
		$a_s$ (10$^6$ m) & 237.95 & 527.11 & 435.91 & 583.52 \\
		$e_s$ & 0.0047 & 0.0013 & 0.0011 & 0.0014 \\
		$i_s$ (deg) & 26.74 & 27.08 & 98.11 & 97.83 \\
		\hline
		\end{tabular}
    	\label{tab:fidmodmoons}
	\end{center}
\end{table*}

We used as fiducial models those small to mid-sized icy Solar System moons most likely to harbor subsurface oceans but (presumably) without a molten interior: Enceladus, Rhea, Titania and Oberon. Table \ref{tab:fidmodmoons} lists all relevant host planet and moon parameters; in all fiducial models the host star is equal to the Sun, namely mass 1 $M_\odot$, radius 1 $R_\odot$, and luminosity 1 $L_\odot$. \\
\indent Least certain are the tidal efficiency factor $\Phi_s$, the regolith cover thickness $D_{rego}$, and the NH$_3$ mass fraction $f_{m,NH_3}$. $\Phi_s$ has been fairly well constrained for a few bodies: \cite{Nimmo2018} find that $0.0026\leq \Phi_s\leq 0.0127$ for Enceladus, so we took as fiducial value $\Phi_s=0.005$. They also cite $\Phi_s\approx 0.015$ for Io, implying an increase in $\Phi$ with increasing mass (although there is no reason to assume this correlation to be causal); since Rhea, Titania and Oberon are considerably more massive than Enceladus but also (since they are farther from their hosts, allowing for less tidal heating and thus less partial melting of the interior) presumably less deformable, we adopt $\Phi_s=0.01$ for the other three moons. \\
\indent The regolith cover thickness is unknown for any body beyond our own moon (between 1 and $\sim$20 m, with similar conditions expected on Mercury; see \citealt{ShkurBonda2001} and \citealt{YuFa2016}) and Mars (based on impact craters, estimates are $\sim$100 m) so we must make an informed estimate. Since Enceladus' tidally induced cryovolcanism creates the icy E Ring around Saturn \citep{Kempf2018} this material is scooped back up by both Enceladus itself and the other moons of Saturn. In addition, since these moons are all less massive than Luna, their gravity compresses the porous regolith less. As a result, we expect the regolith cover on Rhea to be thicker than on Luna: we adopt 50 m. The Uranian system is thought to have a similar, tidally active past \citep{Desch2007} so we also assume 50 m for Titania and Oberon. Conversely, Enceladus is still tidally active and shows signs of recent resurfacing \citep{Nimmo2018} so here we assume the regolith cover to be thinner; since the surface is so young, regolith has not yet had the time to pile up. We assume 20 m. \\
\indent The NH$_3$ mass fraction is known to be greater in the outer Solar System, where conditions allow for the condensation of NH$_3$ ices into the accreting moons. More generally, $f_{m,NH_3}$ goes up as temperature (and thus, illumination) goes down. Suggestions of Enceladus' recent in situ formation as opposed to from Saturn's circumplanetary disk 4.5 Gyr ago \citep{Truong2019,Glein2018} however makes it unlikely to possess extensive NH$_3$ deposits. Therefore, we adopt a no NH$_3$ model for Enceladus, 10\% for Rhea because it is farther from Saturn and, possibly, indigenous to the Saturnian system, and 15\% for Titania and Oberon since they are farther from the Sun.

\subsubsection{Model assumptions}
\label{sssec:appr/meldep/modass}
We assume that our moons are rounded, small (less than $10^{22}$ kg, i.e., about half the mass of Triton) and composed of primarily ices with an NH$_3$ mass fraction between 0 and 15\%, plus possibly silicates or metals. We also assume they possess uniform interiors beneath their ocean layer and that their crust consists of a thick ice shelf topped by a relatively thin layer of regolith (see Figure \ref{fig:intstruc} for our assumed internal structure). The ice crust is assumed to constitute at most 30\% of the moon's radius and at least 100 m to support the top layer of regolith. We also assume that the host planet has a constant temperature on both the day and night sides, based on the rapid rotation and dense atmospheres found on Solar System giant planets, as well as some exoplanets (e.g., \citealt{Snellen2014}). \\
\indent Because of our phenomenological approach to tidal heating, which treats the body as one uniform object, we can not properly account for molten interiors; hence, familiar icy moons such as Callisto and Ganymede fall beyond our reach. Enceladus and the other major Saturnian and Uranian moons are prime examples of the type of moon we are interested in.

\section{Results}
\label{sec:results}
We use four Solar System satellites as fiducial objects: Enceladus, Rhea, Titania and Oberon. For each fiducial we vary the eight most important parameters of our melting depth parameters within the following ranges:

\begin{itemize}
    \item $10^{-2} L_\odot \leq L_* \leq 10^{2} L_\odot$;
    \item $10^{25}$ kg $\leq M_p \leq 10^{28}$ kg;
    \item $1$ AU $\leq a_p \leq 10^{2}$ AU;
    \item $10^{19}$ kg $\leq M_s \leq 10^{22}$ kg;
    \item $10^{8}$ m $\leq a_s \leq 10^{9}$ m;
    \item $10^{-4} \leq e_s \leq 10^{-1}$;
    \item $1$ m $\leq D_{rego} \leq 10^{2}$ m.
    \item $10^{-4} \leq \Phi \leq 10^{-1}$
\end{itemize}

For each pair of parameters, a 2D colormap plot is generated where different shaded regions indicate different temperature and thus melting depth regimes. The full plots are shown in Appendix \ref{app:fullmdgrids}; this section highlights several key subplots. \\
\indent Figures \ref{fig:medeEnc}, \ref{fig:medeRhea}, \ref{fig:medeTit} and \ref{fig:medeOber} show the full melting depth plot grids for Enceladus, Rhea, Titania and Oberon respectively, showing melting depth in terms of their radii (labeled $R_s$ on the colorbars). All parameters have been plotted logarithmically. The gradient region indicates where an ocean might exist between 100 m and 30\% of the moon's radius. Red regions indicate where the surface temperature is above the ice sublimation temperature (i.e., 150 K); dark gray regions indicate a melting depth greater than $0.3R_s$; and light gray regions indicate a melting depth less than 100 m, or where the temperature underneath the regolith cover is already above the melting temperature. The white or black crosses (color chosen for maximum visibility) mark the fiducial values for the tested moon. We note that for the Uranian moons, the red regions extend notably further out from their host star owing to their dark regolith cover ($\alpha_{Titania}=0.17$, $\alpha_{Oberon}=0.14$); this absorbs the exogenic flux more effectively than Enceladus' and Rhea's ice and snow cover.

\subsection{Enceladus}
\label{ssec:results/Enceladus}

\begin{figure*}
    \includegraphics[width=1.05\columnwidth]{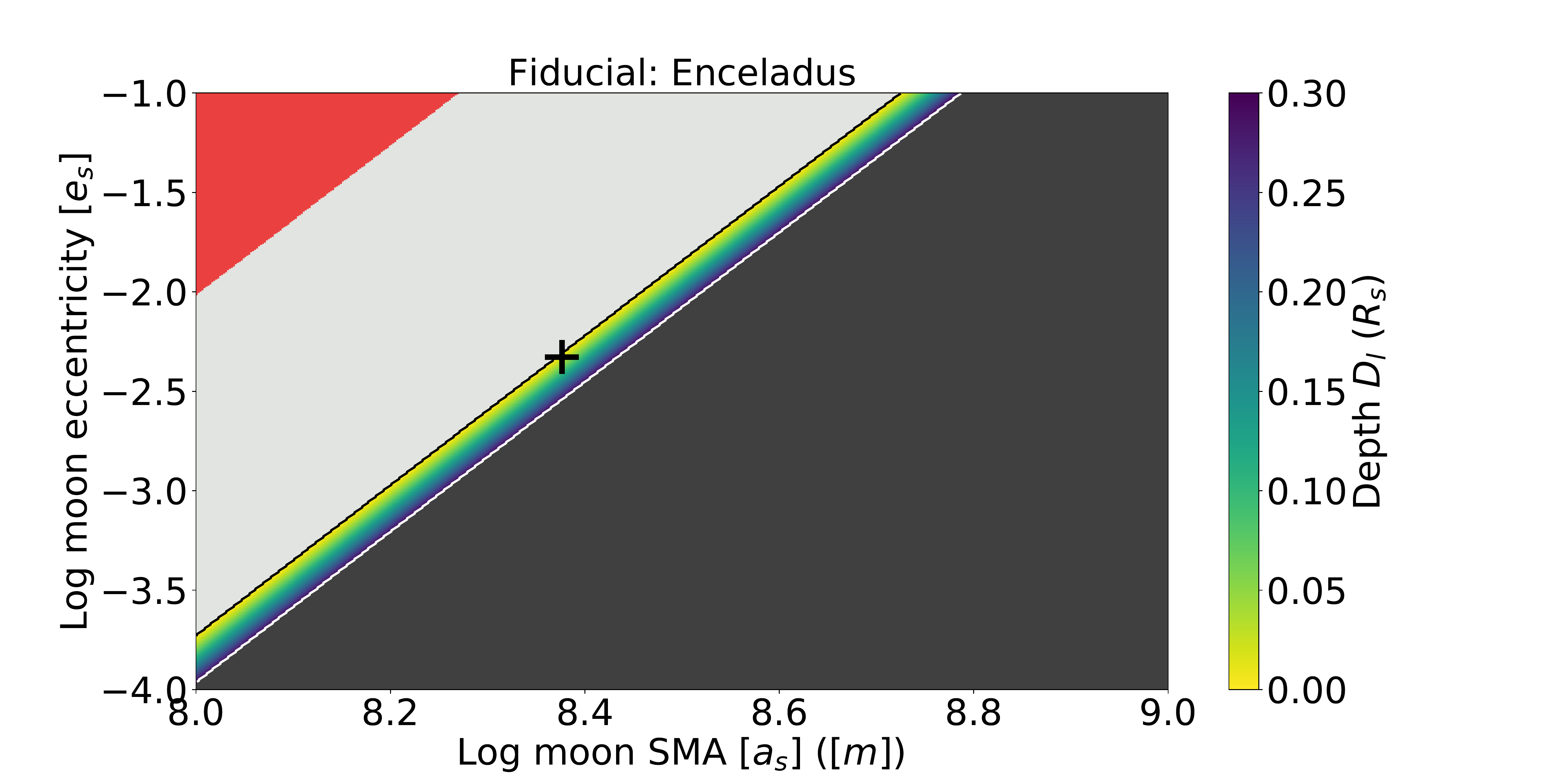}
    \includegraphics[width=1.05\columnwidth]{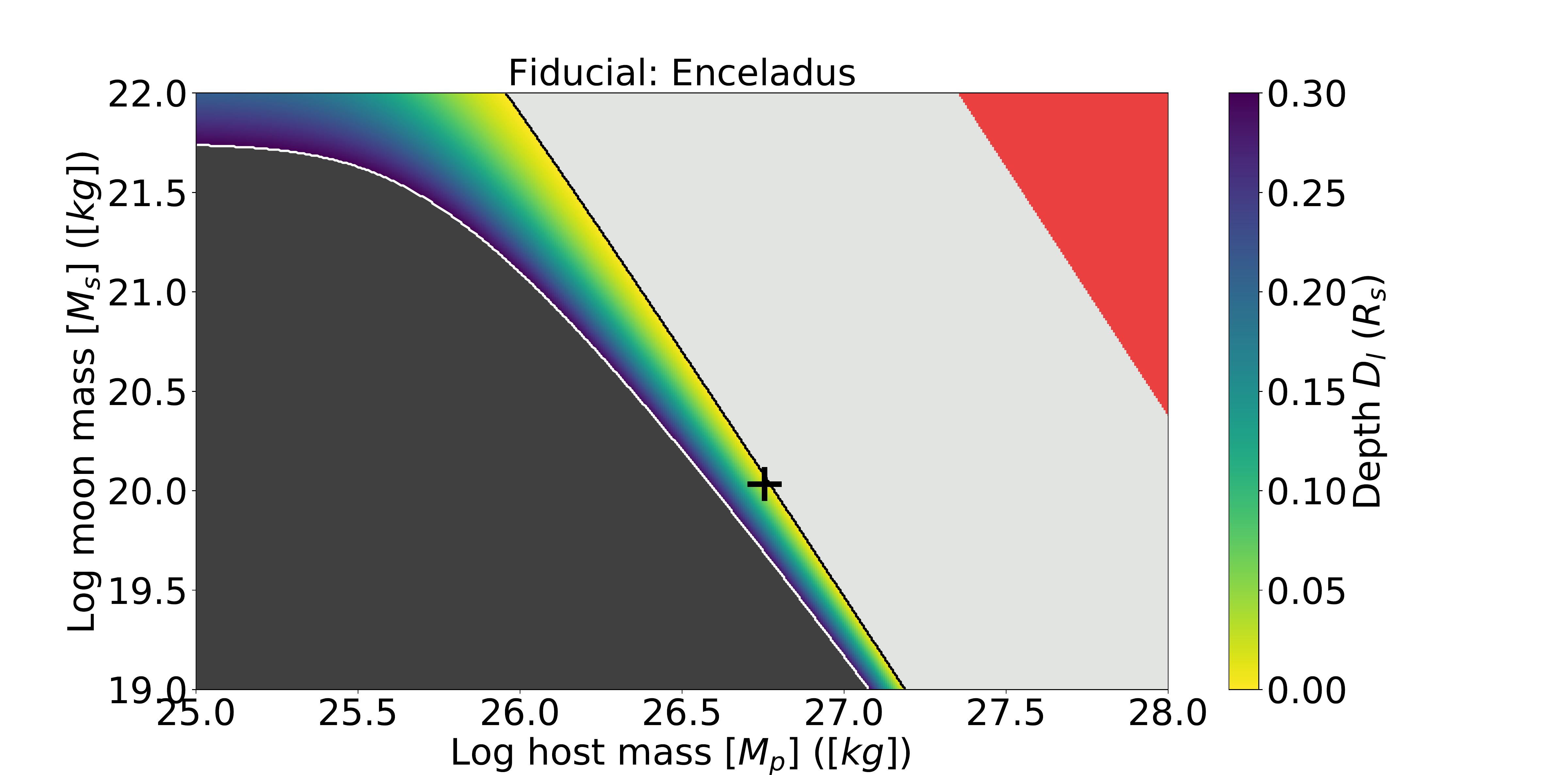}
    \includegraphics[width=1.05\columnwidth]{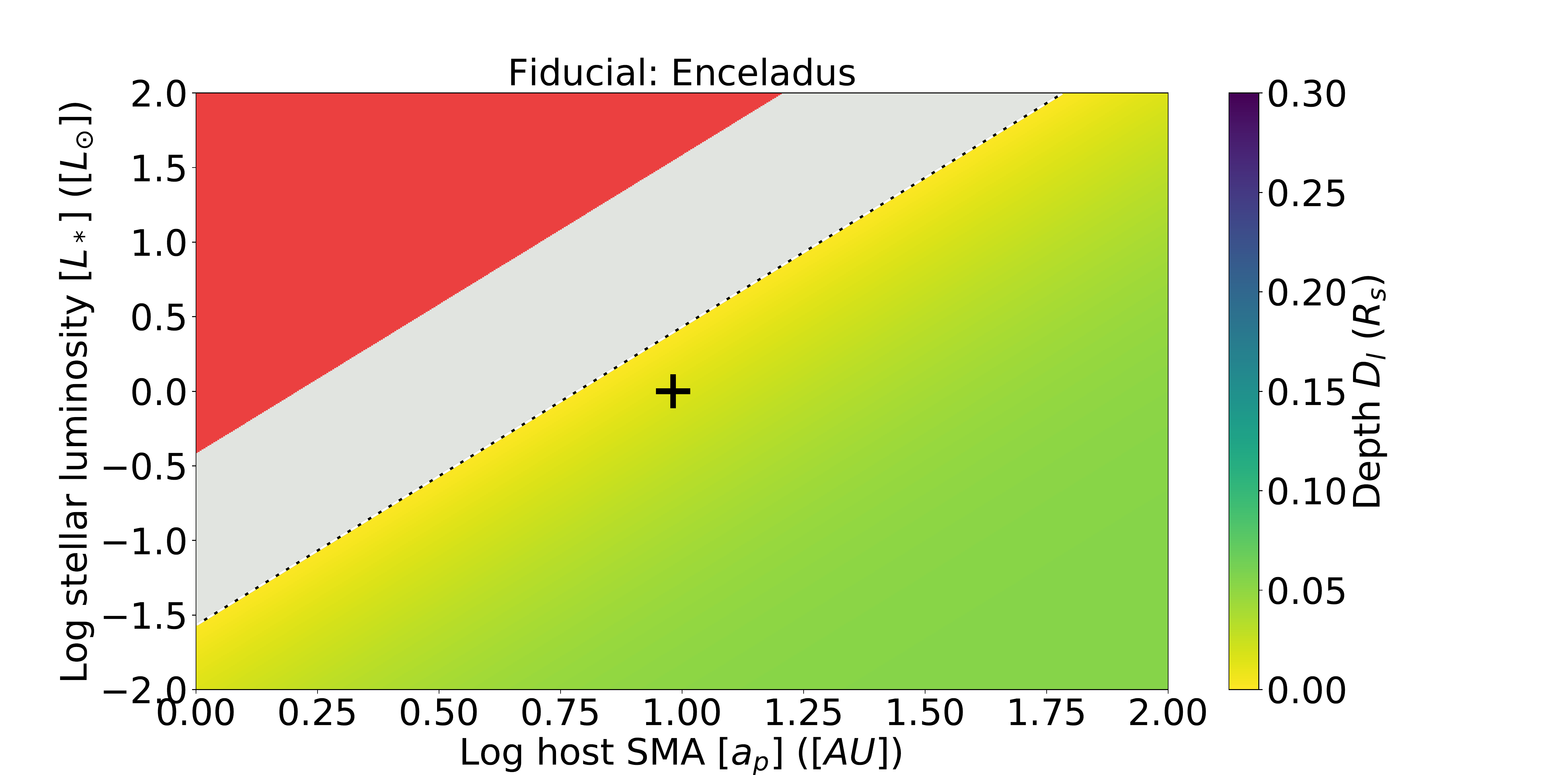}
    \includegraphics[width=1.05\columnwidth]{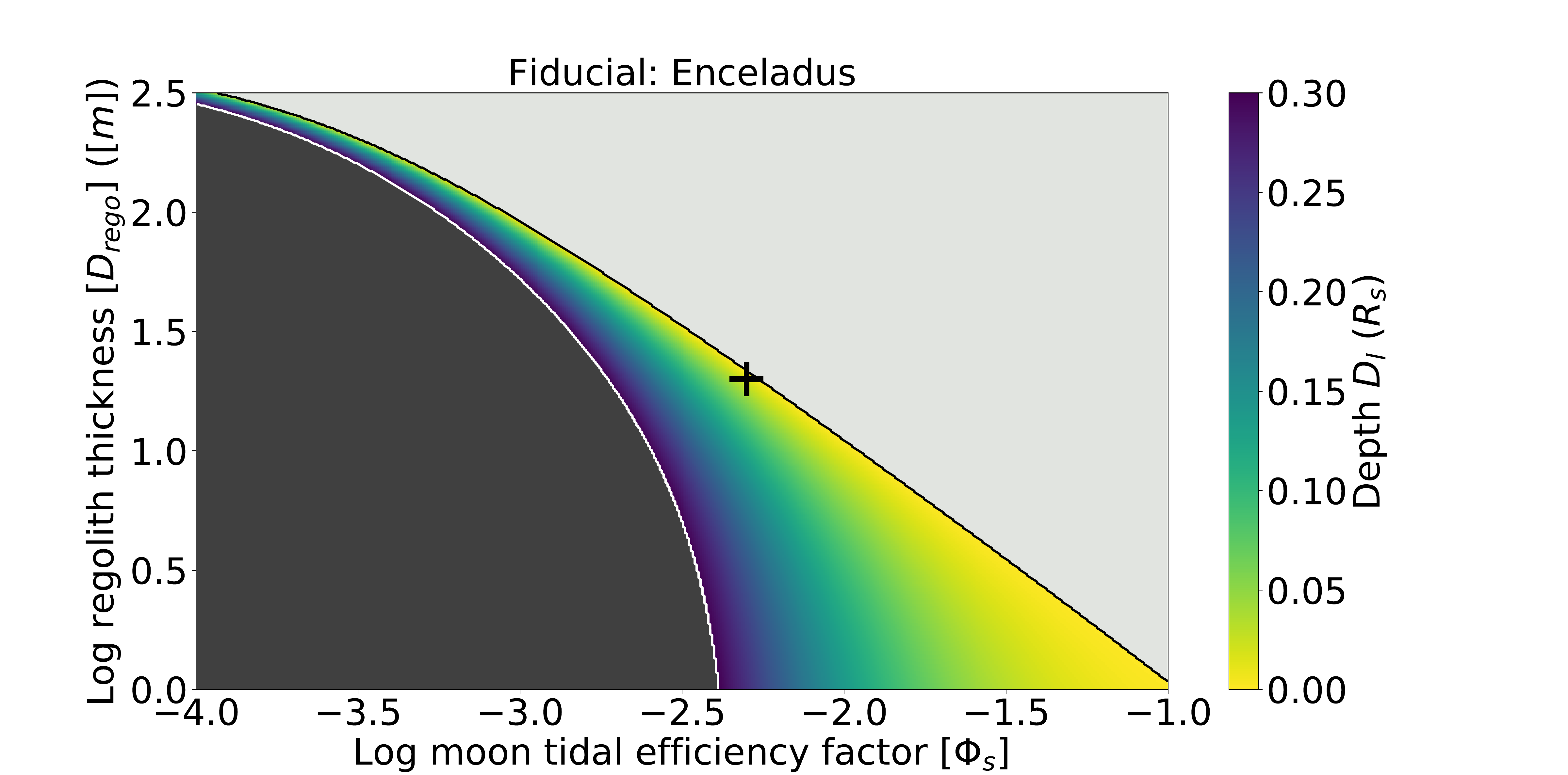}
    \caption{Enceladus' melting depth in terms of satellite radius ($R_s$) as a function of orbital characteristics (top left), planet-moon masses (top right), host semi-major axis (SMA in labels) versus stellar luminosity (bottom left) and moon tidal efficiency versus regolith thickness (bottom right). Red regions mean a desiccated moon, dark gray a frozen moon, and light gray one with at most a thin ice film; crosses represent fiducial values. For Enceladus itself, our model predicts a melting depth of 3.6 km.}
    \label{fig:singleEnc}
\end{figure*}

\begin{figure}
    \includegraphics[width=1.05\columnwidth]{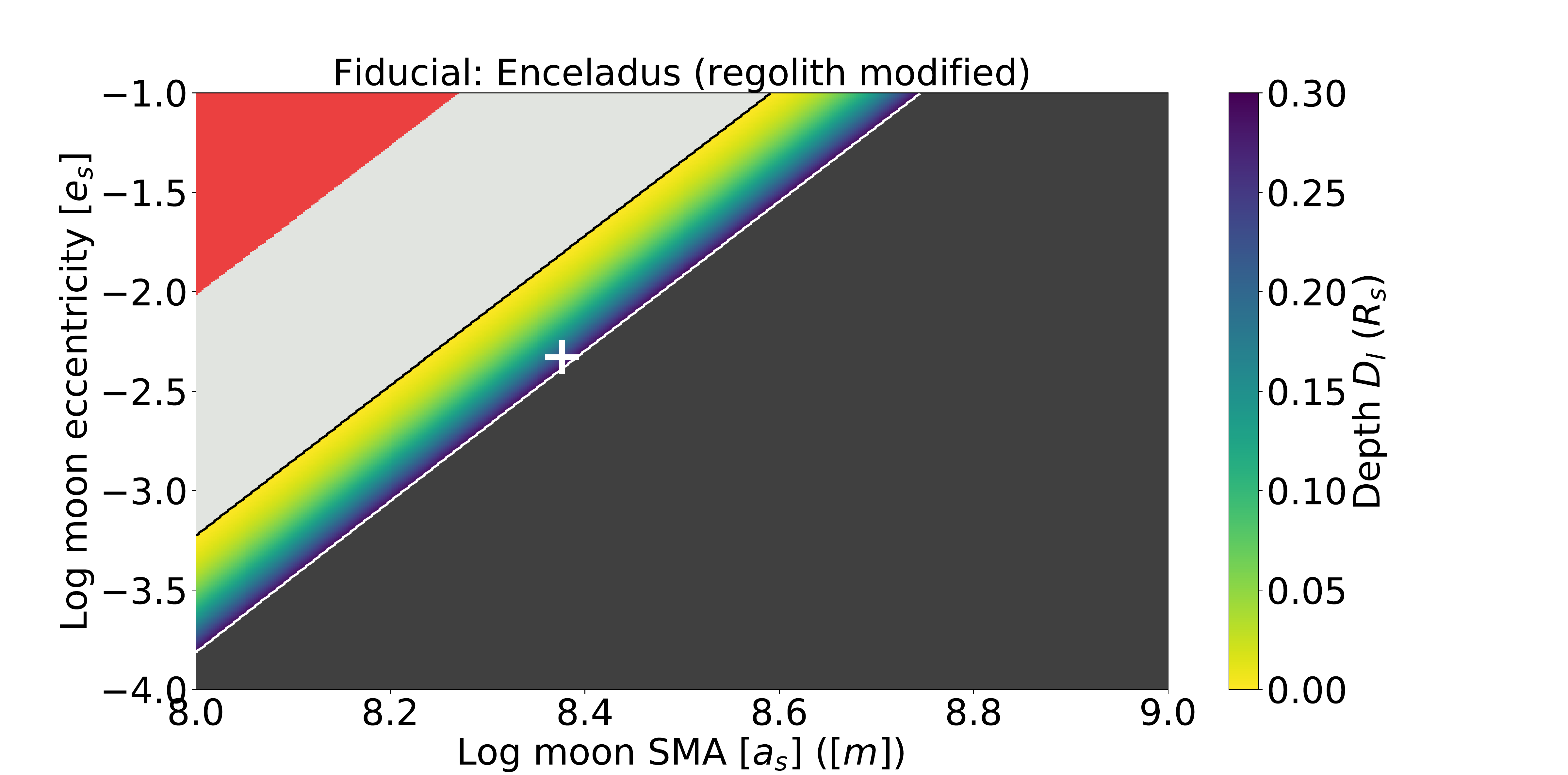}
    \caption{Same as previous figure, top left panel, but for a ten times higher regolith thermal conductivity: 0.01 W m$^{-1}$ K$^{-1}$ as opposed to the 0.001 W m$^{-1}$ K$^{-1}$ used in all other figures. The range of parameters allowing for a subsurface ocean is now noticeably wider, and our estimate for the melting depth is considerably deeper ($\sim$50 km).}
    \label{fig:EncRegoMod}
\end{figure}

Our model predicts a subsurface ocean at a depth of 3.6 km. This is shallow compared to the 10-30 km conventionally cited in literature \citep{Nimmo2018}. It must be noted that firstly, our model assumes a radially symmetric moon and secondly, it does not take into account cryovolcanic activity: it is purely conductive. More detailed models of Enceladus invoke an asymmetric structure wherein the south polar crust is considerably thinner (as thin as 2 km versus $\sim$25 km for the rest of the moon; see \citealt{Cadek2016,Beuthe2016,Glein2018}). \\
\indent Figure \ref{fig:singleEnc} highlights several key results from the plot grid of Figure \ref{fig:medeEnc}. It is evident from the top left panel that Enceladus' semi-major axis and eccentricity most strongly influence its condition: slight variation from observed values rapidly leads to it becoming either frozen or desiccated. The range of possible orbital characteristics allowing for a subsurface ocean is very narrow. Similarly, plotting host mass versus moon mass (top right) shows that a hypothetical exomoon on Enceladus' orbit would be desiccated quickly given a more massive host. Conversely, putting host semi-major axis against stellar luminosity (bottom left) shows that independent of how faint the host star might become or how distant the host planet might orbit, an Enceladus-like moon around a Saturn-like host would retain its subsurface ocean. This supports the idea that Enceladus' liquidity is maintained primarily by tidal heating, although it must be noted that our value for $\Phi$ was informed by papers discussing a potential Enceladean subsurface ocean to begin with: as shown by the lower right panel, choosing another tidal efficiency factor or regolith thickness can easily either freeze or desiccate Enceladus. \\
\indent The tightness of the range of orbital characteristics allowing for a subsurface ocean seems curious, but it is no cause for concern and may be attributed to various model assumptions. Changing the fiducial regolith thickness or thermal conductivity can affect the width of the ocean regime (see Figure \ref{fig:EncRegoMod}). Similarly, our ignoring of volcanism deprives Enceladus of an additional pathway to release internal heat; if one takes volcanism into account the moon's internal temperature could drop more efficiently (which is effectively equivalent to increasing the regolith conductivity) and the ocean regimes in Figure \ref{fig:singleEnc} may widen.

\subsection{Rhea}
\label{ssec:results/Rhea}

\begin{figure*}
    \includegraphics[width=1.05\columnwidth]{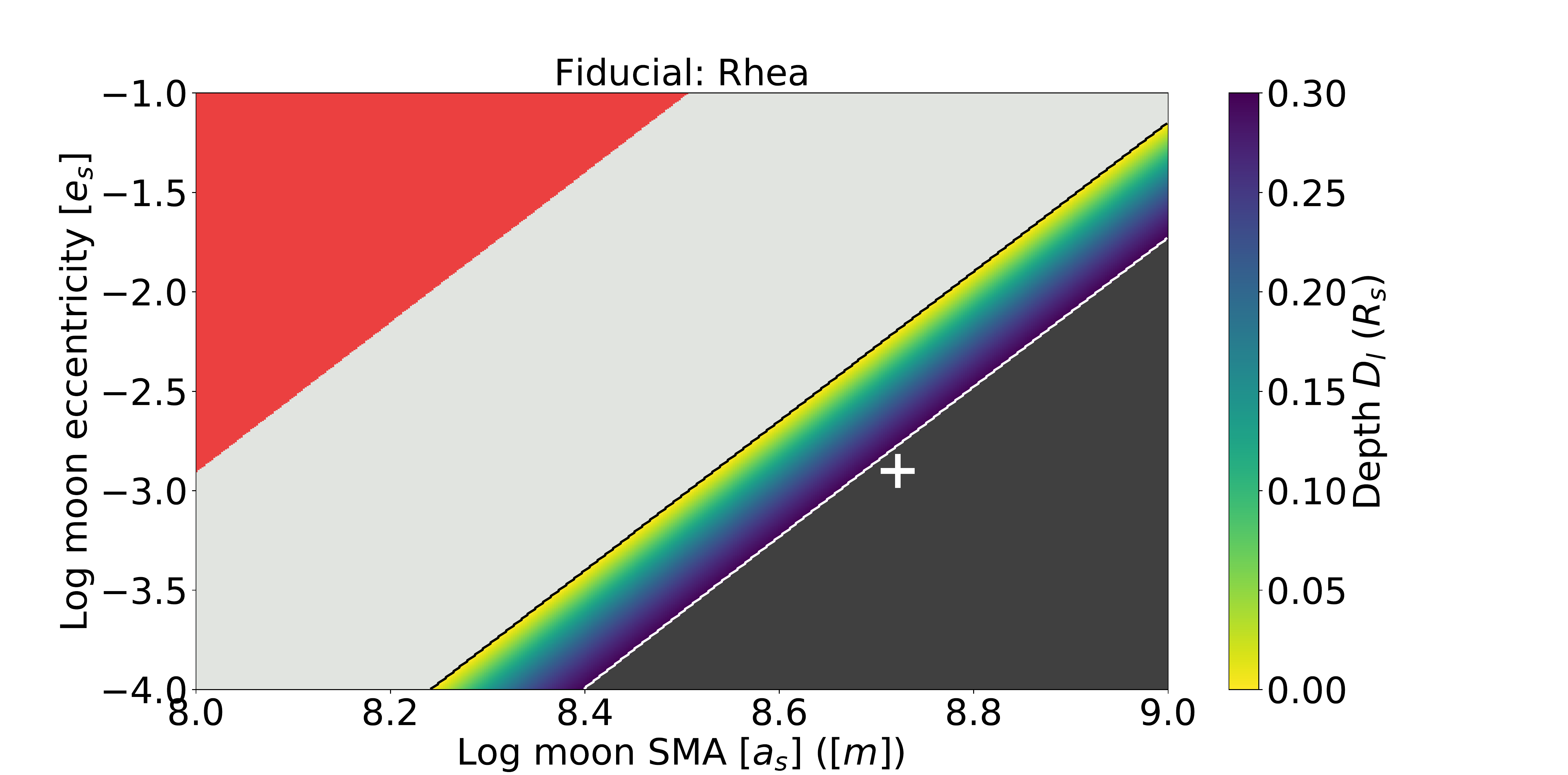}
    \includegraphics[width=1.05\columnwidth]{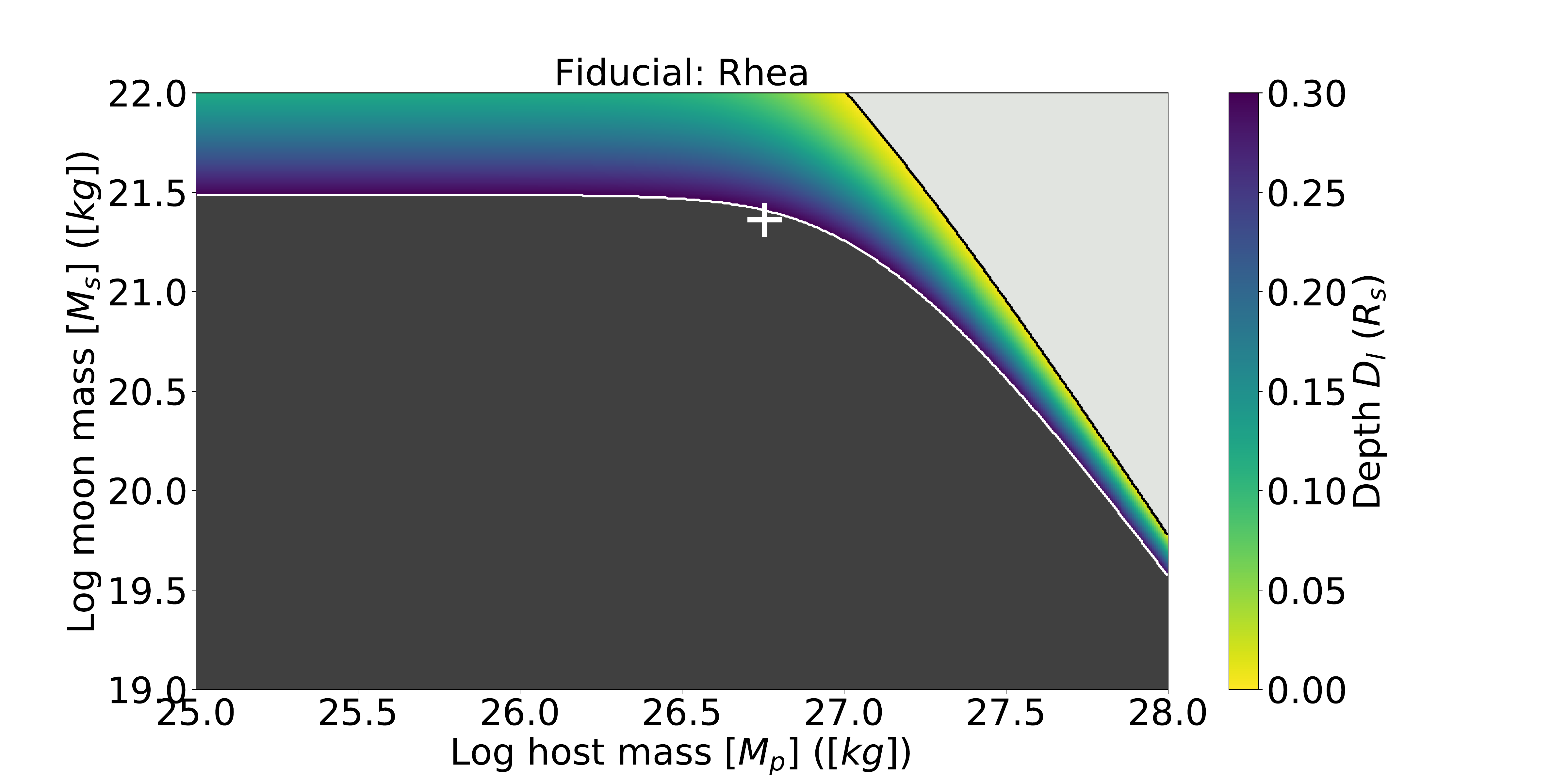}
    \includegraphics[width=1.05\columnwidth]{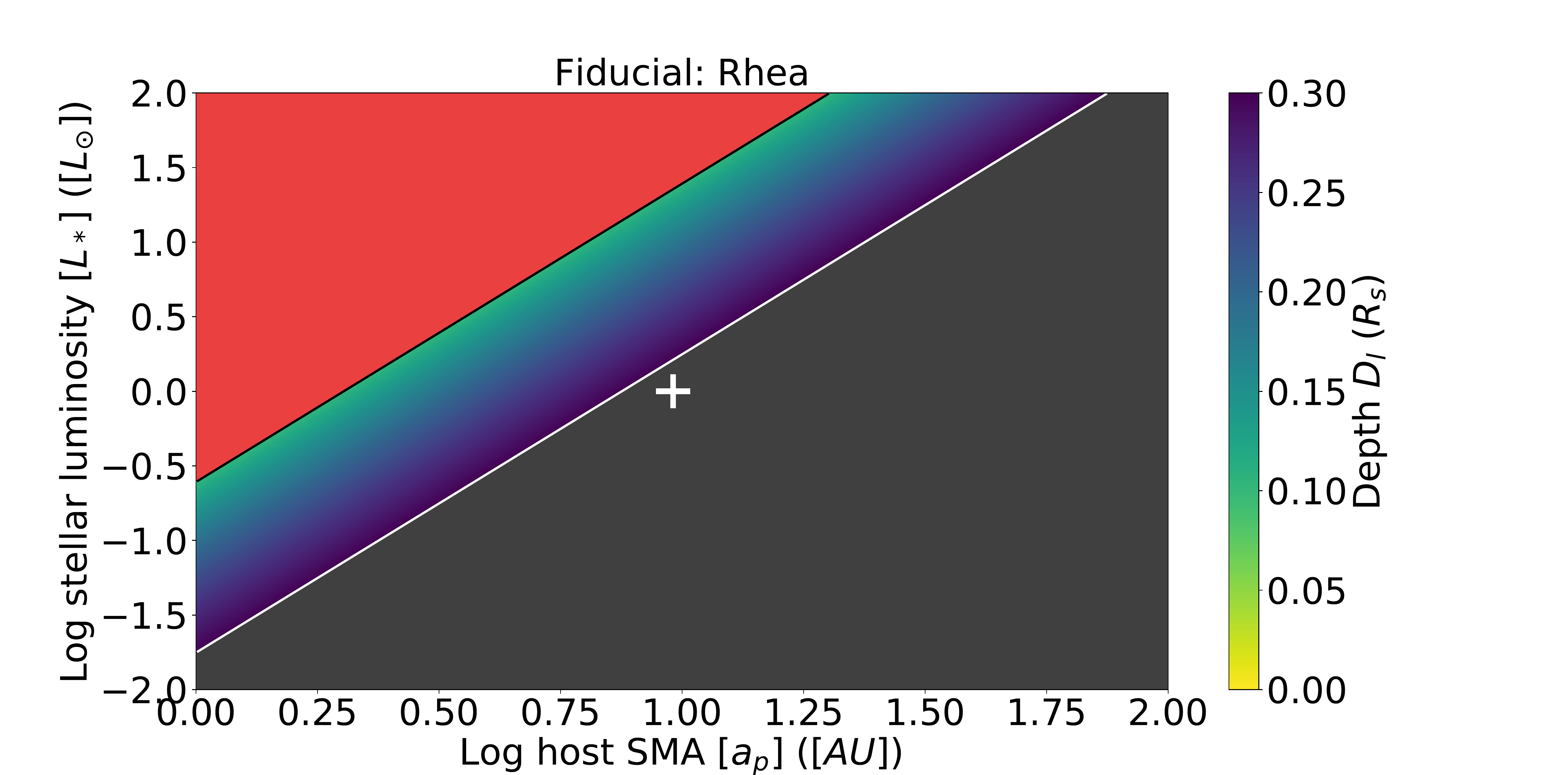}
    \includegraphics[width=1.05\columnwidth]{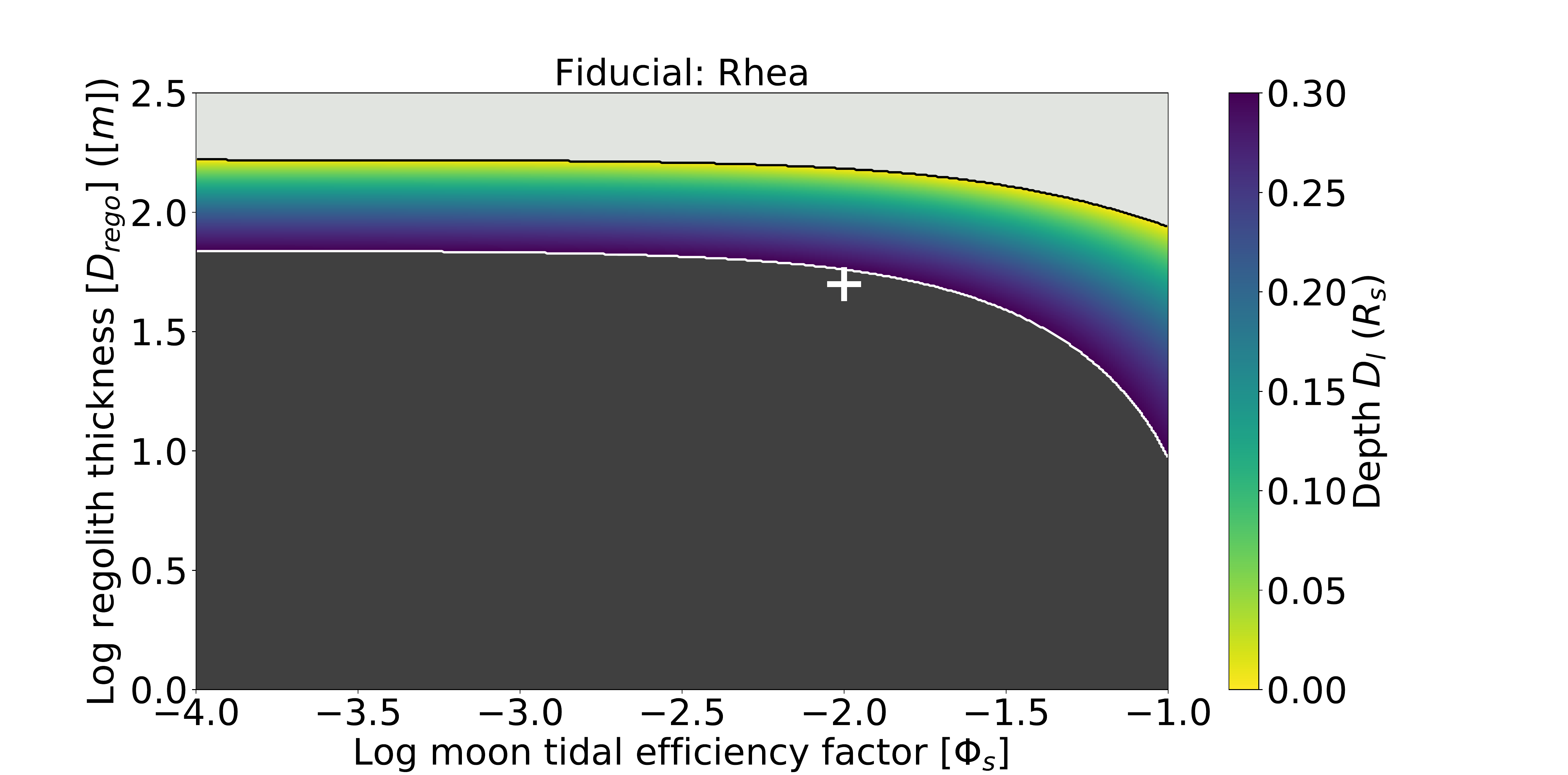}
    \caption{As Figure \ref{fig:singleEnc}, but for Rhea. Our model predicts a fully frozen body.}
    \label{fig:singleRhea}
\end{figure*}

Figure \ref{fig:singleRhea} highlights the same parameters as Figure \ref{fig:singleEnc}. We see that Rhea seems unable to maintain an ocean above 30\% of its radius. Since it receives the same exogenic flux (slightly less, since the planetary contribution is lower; the stellar component is identical, however) as Enceladus, that our model fails to predict an ocean is presumably because of weaker tidal forces from Saturn. Rhea is not massive enough, not close enough to Saturn, or not eccentric enough to maintain sufficient levels of internal heat; changing any of these three parameters can successfully push the melting depth across our adopted 30\% moon radius melting threshold. \\
\indent Owing to Rhea's larger mass, the range of orbital parameters allowing for a subsurface ocean is slightly wider (top left) than for Enceladus. A mildly more eccentric orbit (0.002 rather than the observed 0.0013) would have allowed Rhea to sustain an ocean via tidal heating; similarly, a more massive moon (top right; $30\cdot 10^{20}$ kg rather than the observed $23.06\cdot 10^{20}$ kg) would have been able to sustain one via residual endogenic processes (given our method of computing such heat, i.e., Equation \ref{eq:radaccheat}). We also see that exogenic processes could have made a difference for Rhea by raising the surface equilibrium temperature (bottom left); a brighter host star or a closer host planet orbit would have allowed for a subsurface habitable environment. This shows that moons with insufficient tidal heat must gather the requisite heat from different sources: increasing Rhea's tidal efficiency factor (bottom right) by half an order of magnitude or thickening its regolith blanket would see it melt into an Enceladus-like state.

\subsection{Titania \& Oberon}
\label{ssec:results/Titaberon}

\begin{figure*}
    \includegraphics[width=1.05\columnwidth]{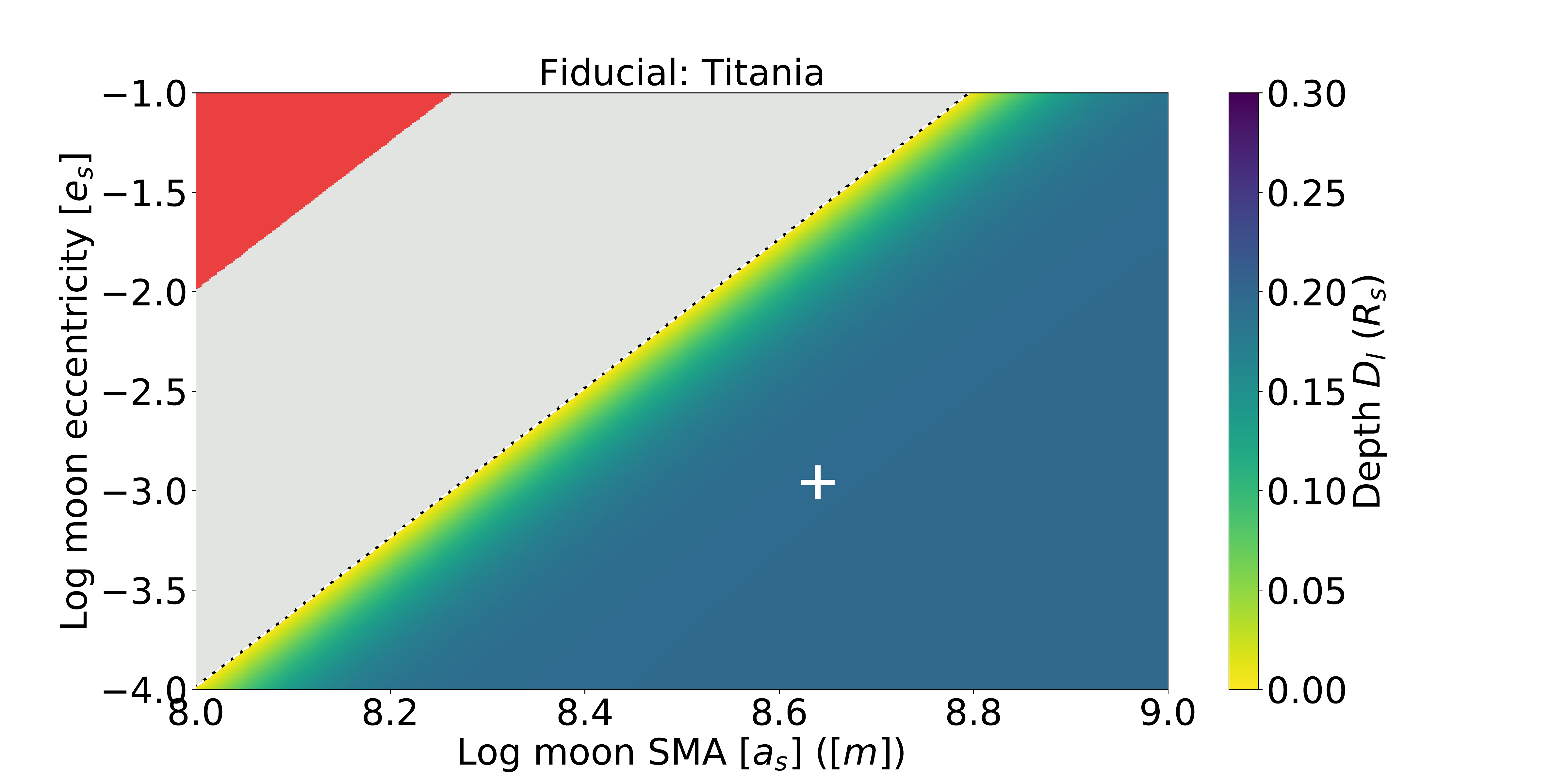}
    \includegraphics[width=1.05\columnwidth]{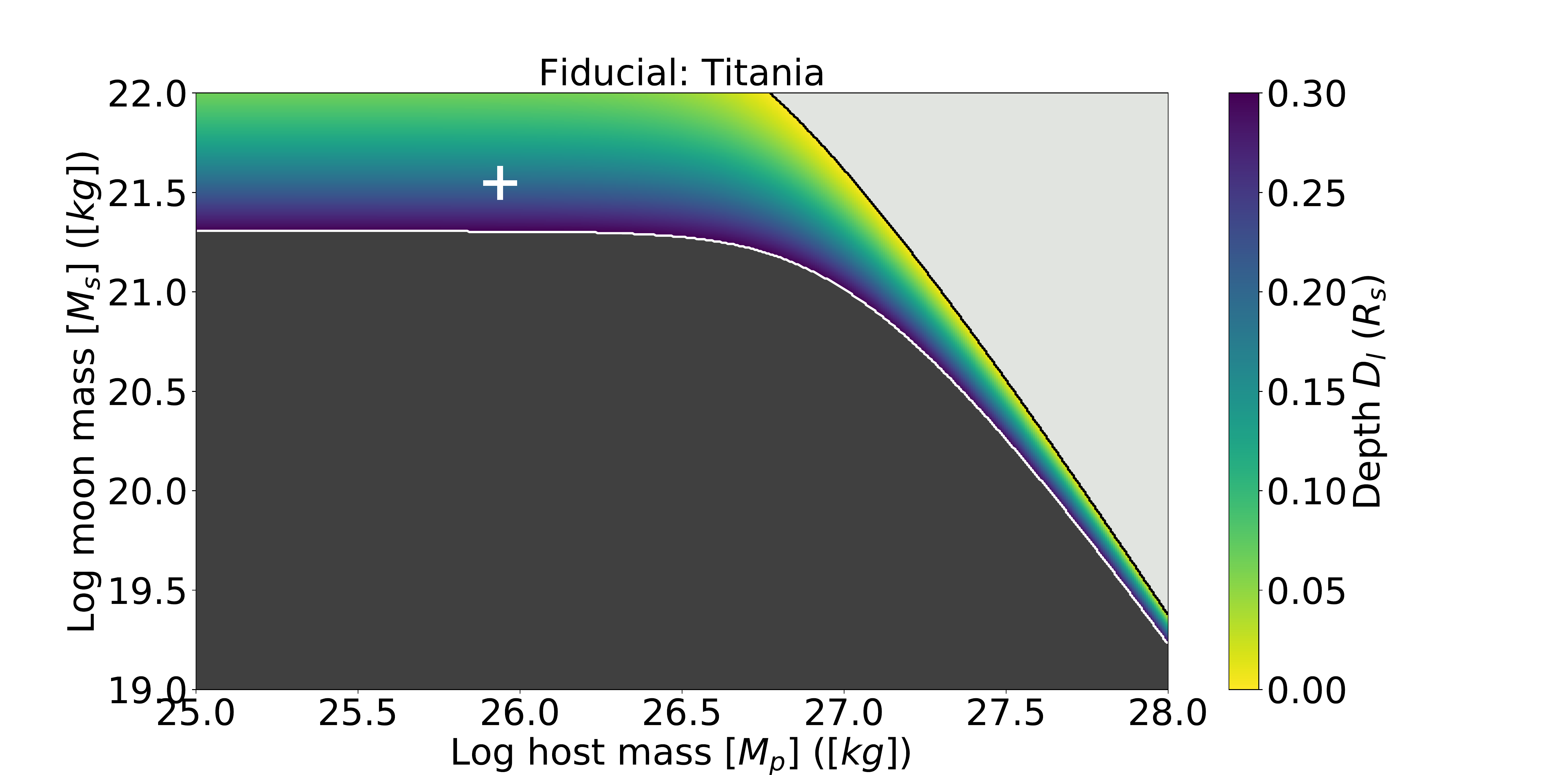}
    \includegraphics[width=1.05\columnwidth]{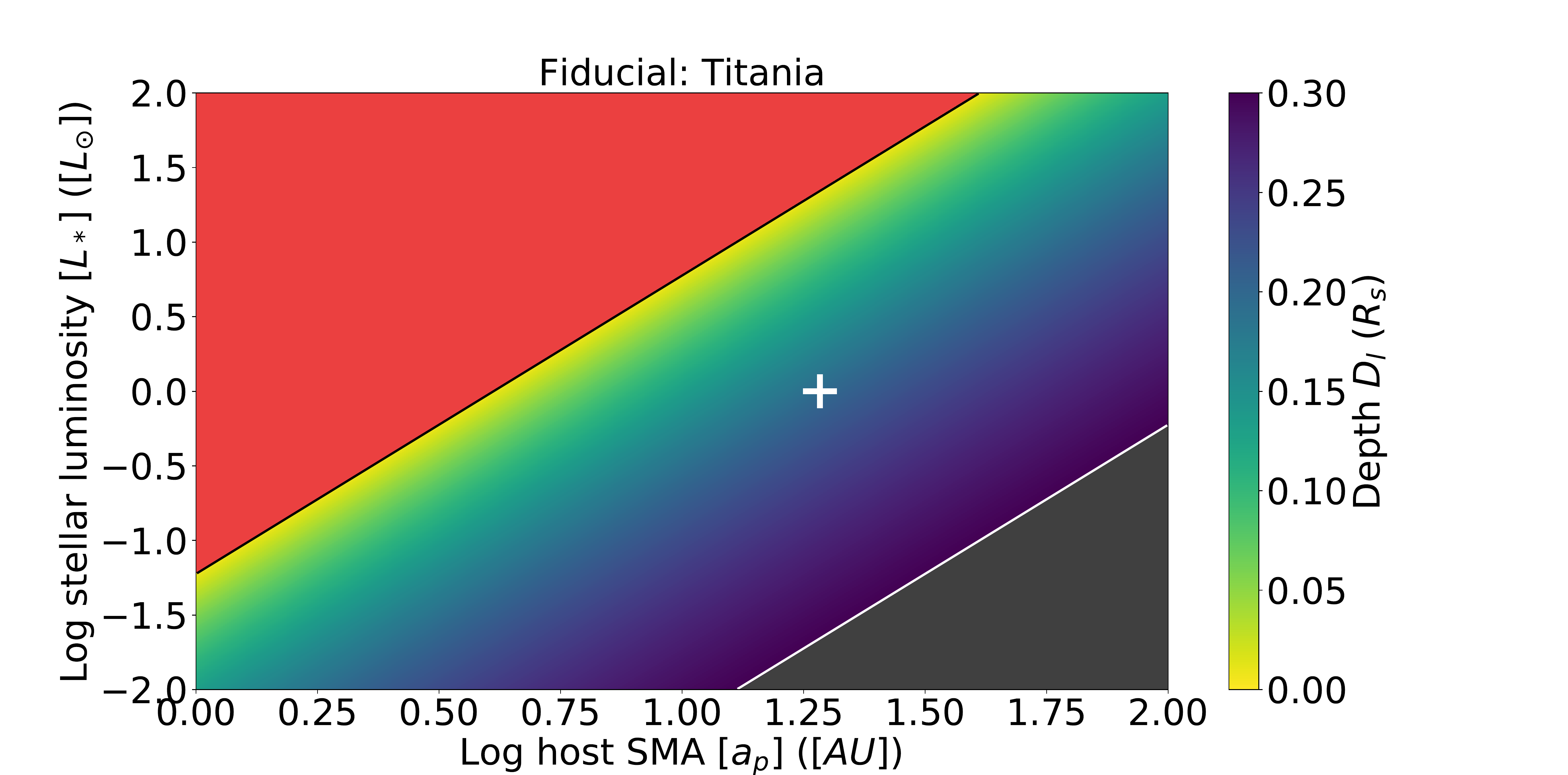}
    \includegraphics[width=1.05\columnwidth]{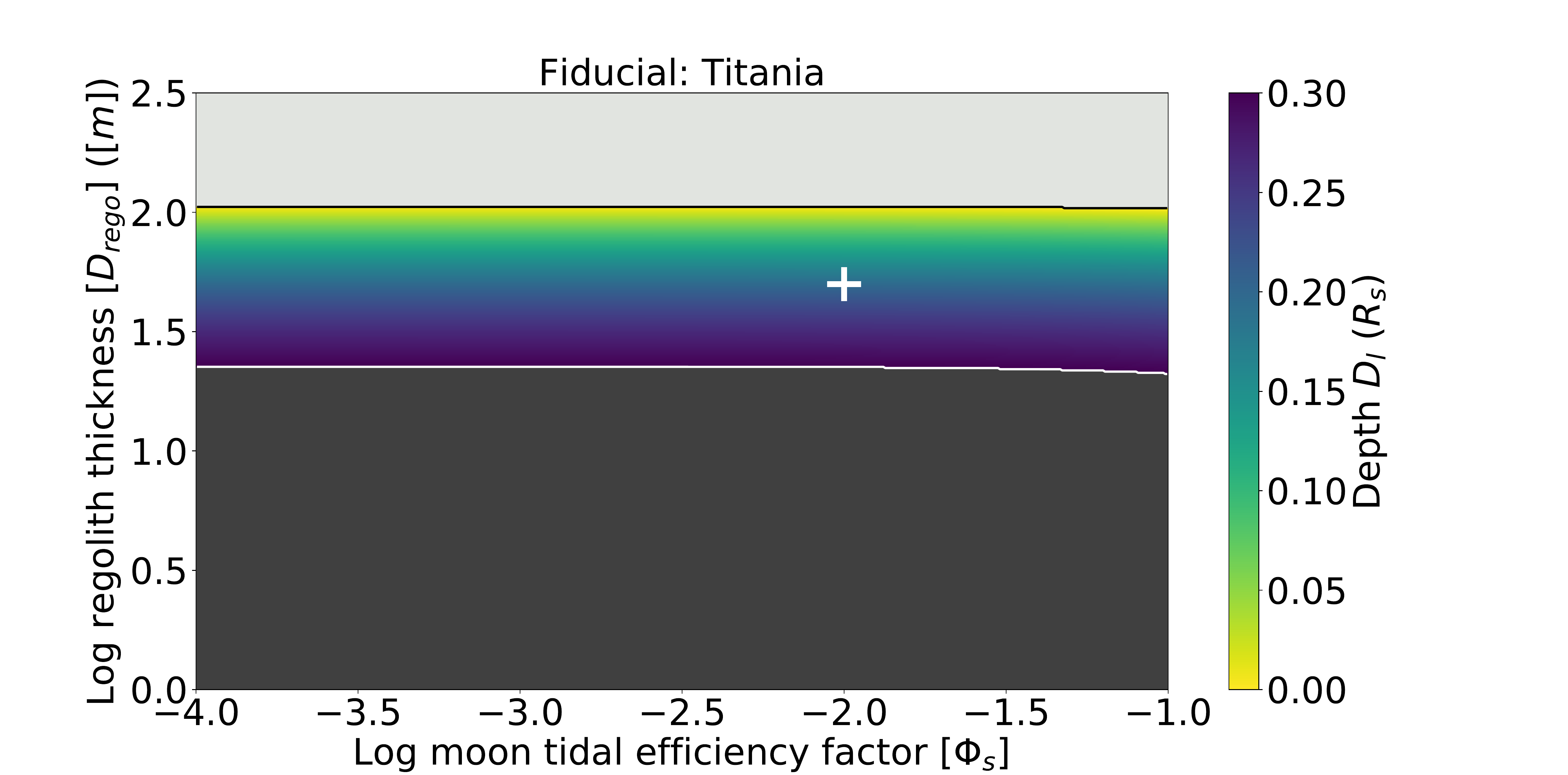}
    \caption{Same as previous figure, but for Titania. Our model predicts a melting depth of 154 km.}
    \label{fig:singleTit}
\end{figure*}

\begin{figure}
    \includegraphics[width=1.05\columnwidth]{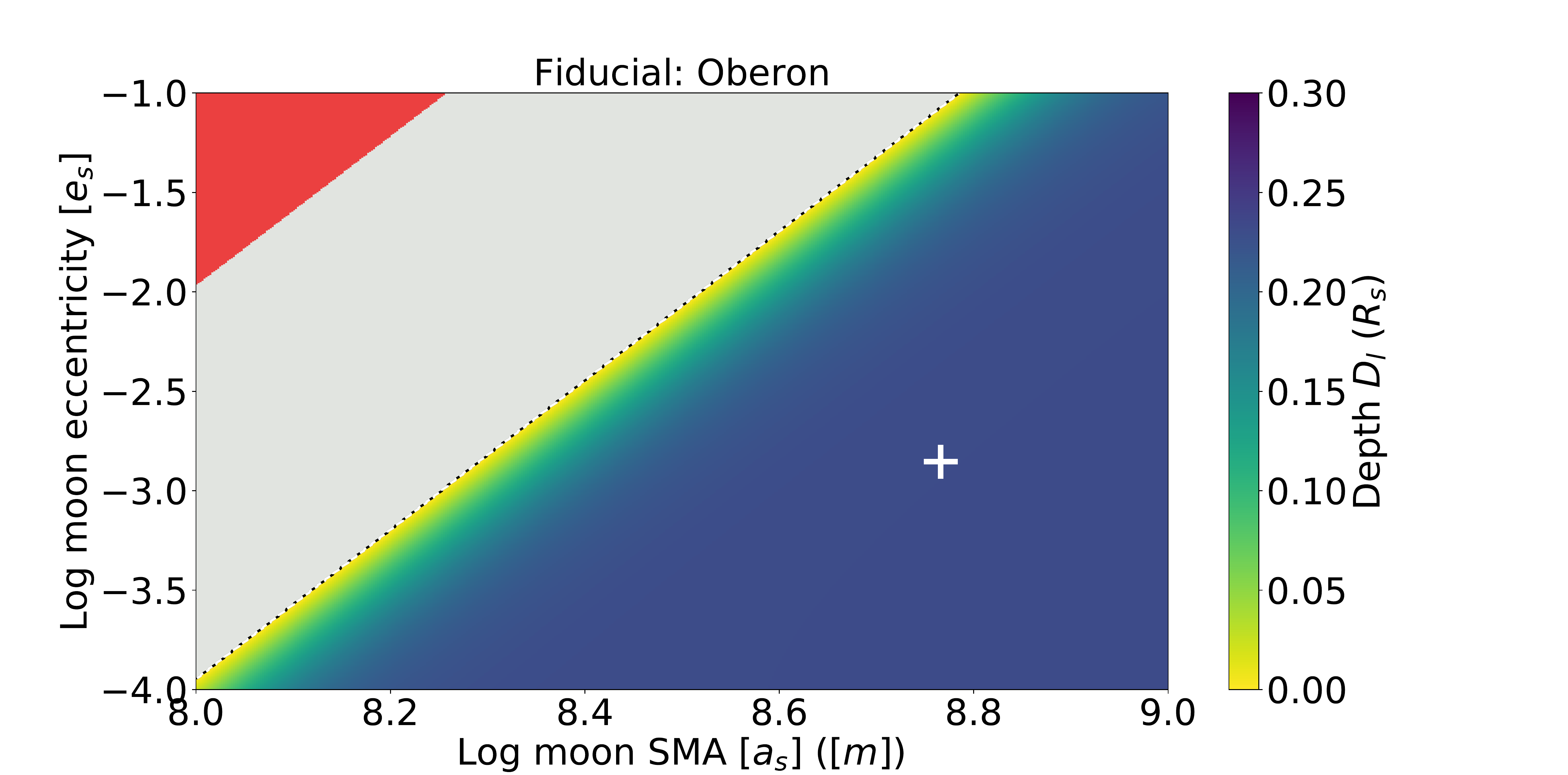}
    \caption{Same as previous figure, but for Oberon, and only for the orbital characteristics (since Oberon's mass and tidal efficiency is similar to Titania's and both Uranus' and the Sun's characteristics are the same for all Uranian moons). Our model predicts a melting depth of 176 km.}
    \label{fig:singleOb}
\end{figure}

Figure \ref{fig:singleTit} again highlights the same parameters as Figure \ref{fig:singleEnc}; Figure \ref{fig:singleOb} shows only the orbital characteristics-subplot for Oberon, since the other two panels would be very similar to Titania's. Both figures show that Titania and Oberon would maintain their internal oceans irrespective of lower eccentricity or larger semi-major axis (top left), implying they are not primarily dependent on tidal heat. This is further supported by the bottom right panel, which shows that a change in the tidal efficiency factor would barely influence the melting depth. Since exogenic flux at Uranus' semi-major axis is negligible, this implies radiogenic and formation heat alone may be responsible for sustaining their subsurface oceans. They are also aided by their high mass (top right), thick crusts, dark surfaces, and high NH$_3$ fractions. The Titanian and Oberonian oceans are however situated at far greater depth than Enceladus': below 150 km, or over 20\% of their radii. The Uranian moons also respond more intensely to exogenic heating because of their dark surfaces (bottom left): were Uranus closer than $\sim$4 AU to the Sun, their ice crusts would sublimate.

\section{Discussion}
\label{sec:disc}

\subsection{Plausibility}
\label{ssec:disc/plausb}
Our estimate for the thickness of Enceladus's ice crust is, at 3.6 km, smaller than literature values of 10-30 km \citep{Nimmo2018}. This may be a consequence of our model being purely conductive: we do not consider the effects of cryovolcanism nor ice shell convection. Volcanic activity may serve as an additional pathway to release excess internal heat, thus lowering the internal temperature and leading to a thicker ice crust. We also see in Figures \ref{fig:singleEnc} and \ref{fig:singleRhea} that there seems to exist only a narrow band of possible semi-major axis-eccentricity states for which primarily tidally heated moons (i.e., Enceladus and Rhea) may sustain subsurface oceans; at large semi-major axis, extreme eccentricities are required to keep the moon liquid, begging the question of whether such configurations are in fact physically possible. It is however important to note that the tightness of this ``liquid band" can change if different values of the regolith thickness or thermal conductivity are chosen (see Figure \ref{fig:EncRegoMod}), which is effectively equivalent to adding another outlet of internal heat; hence, our ignoring alternative heat outlets may influence our results. \\
\indent Titania and Oberon are both not known to be volcanically active but in their cases our knowledge regarding internal structure is very limited. \cite{Hussman2006} studied hypothetical subsurface oceans in the outer Solar System, finding ice shell thicknesses well in excess of 200 km. It is important to note that their approach assumes the thermal conductivity of ice takes a single, fixed value. The conductivity of ice is known to change with temperature \citep{AndersInaba2005}; we implemented this gradient throughout the ice shells. $k$ then goes as high as 10 W m$^{-1}$ K$^{-1}$ around 75 K (approximate surface temperature at Saturn's semi-major axis), which could well explain the discrepancy between our results and \cite{Hussman2006}. They also use lower NH$_3$ mixing ratios than in our model, both for Rhea and for Oberon and Titania: our higher fractions of 10-15\% versus their 0.5 to 5\% contributed to our thinner ice crust. Their result for Rhea is different from ours: they predict an ocean at 400 km depth, while our Rhea remained solid. Reevaluation of the Rhea system with different NH$_3$ fractions, tidal efficiency factors, and regolith thickness might result in different predictions. In future studies these three unknowns should be well considered since they can drastically alter results: we adopt conservative values for $\Phi$ and $D_{rego}$, but it is well possible to melt or freeze an otherwise habitable moon by tweaking these parameters (see Figure \ref{fig:medeDeadRhea} for a desiccated Rhea and Figure \ref{fig:medeFrozenEnc} for a frozen Enceladus).

\subsection{Subsurface habitable zone \& tidal habitable edge}
\label{ssec:disc/subshabzone}
\subsubsection{Circumstellar limits}
\label{ssec:disc/subshabzone/cstellar}
All four fiducial moons rely on some amount of endogenic heating to maintain their subsurface oceans. In Titania's and Oberon's cases this is almost entirely radiogenic and residual formation heat; Enceladus relies on tidal heating, while on Rhea both processes are of similar magnitude. In all cases the influence of exogenic heat is felt predominantly on the surface, but little deeper. This implies that subsurface oceans can be sustained beyond the snow line regardless of distance to the star; the circumstellar subsurface habitable zone for moons may extend up to the edge of the host star's Hill sphere. This finding significantly expands the circumstellar range wherein life-sustaining habitats may be found: considering only terrestrial planets as habitats limits the search to the habitable zone, whereas if one includes icy moons (and, possibly, surface habitable terrestrial exomoons too), potentially habitable environments may be found all across an exoplanetary system, and indeed our own Solar System as well. \\
\indent Several notes must be made. Such an indefinite extent of the habitable zone assumes that the satellite generates or receives enough internal heat to sustain subsurface liquid water. Secondly, if the satellite is small (smaller than Rhea, i.e., $\sim2\cdot10^{21}$ kg), tidal forces become the only viable source of internal heat: any radiogenic heat dissipates at a rate proportional to area over volume, so $\propto R^{-1}$. Thirdly, the moon habitable zone is eventually truncated at large distance from the host star at the edge of the planetary system, since moons require planetary hosts. Lastly, while the circumstellar subsurface habitable zone may extend indefinitely outward, exogenic processes do define an inward bound, namely the point at which the induced surface temperature on the moon exceeds the sublimation point of water ice (around 150 K). In parallel to \cite{HellerBarnes2013}, we could thus regard the snow line as a circumstellar subsurface habitable edge.

\subsubsection{Circumplanetary limits}
\label{ssec:disc/subshabzone/cplanetary}
\indent The circumplanetary habitable zone depends on more factors than the circumstellar habitable zone. The melting depth in total depends on seventeen more or less free parameters. As discussed in Section \ref{ssec:disc/plausb}, the tidal efficiency factor, NH$_3$ mass fraction, and regolith thickness present three major unknowns which are all poorly constrained. Regardless, we can impose some limits on the circumplanetary habitable zone. \\
\indent Firstly, there are dynamical constraints. Both the Roche limit (generally at $\sim 2.5R_p$) and the Hill sphere (at distance $R_H$) limit exomoon orbits, although truly dynamically stable orbits can not exceed $0.5R_H$; all major icy moons of the Solar System orbit well inside that. \\
\indent Secondly, the tidal habitable edge forms another inner limit: this is the closest a moon can orbit without undergoing such tidal heating that it becomes Venus-like, that is, a runaway greenhouse (as described in the introduction; see also \citealt{HellerBarnes2013} and \citealt{ForganDobos2016}). While it is often stated in terms of only moon semi-major axis, our melting depth plot grids show that the habitable edge (the black lines in Figures \ref{fig:medeEnc} to \ref{fig:medeFrozenEnc}) is described by multiple parameters. Prime among these are the moon's semi-major axis, eccentricity, mass, regolith cover, and tidal efficiency factor; however, all seventeen model parameters influence where this edge is located. \\
\indent However, it must be pointed out that the habitable edge was previously described for surface habitability, and so our subsurface variant does not need to coincide with the surface habitable edge. Since it is such a manifold condition, it may in our case be more prudent to speak not of a habitable edge but of a maximum melting criterion: a maximum amount of internal melting allowed before we consider the moon desiccated. Per our model, we find this criterion is satisfied whenever the melting depth as computed by Equation \ref{eq:meltdepthan} is larger than 100 m + $D_{rego}$ (as discussed in Section \ref{ssec:appr/meldep}). This 100 m limit imposed on the ice crust's thickness is arbitrary; another maximum melting criterion may be chosen when more data on icy moon crusts becomes available. Similarly, we can define a minimum melting criterion by arbitrarily picking a maximum fraction of the moon's radius consisting of ice and thus meltable into an ocean. We chose 30\%, but this too may be subject to change if and when more data becomes available. A circumplanetary habitable zone may then be defined between both criteria.

\subsection{Timescales \& implications for fiducial moons}
\label{ssec:disc/impforfidms}
\indent Our analysis only indicates the possibility of liquid water. A key condition for the actual appearance of life is the long-term stability of the environment. The emergence of the first life on Earth required about half a billion years; metazoan life (animals and other complex lifeforms) only appeared over 3 billion years after abiogenesis \citep{Schwieterman2019,Schidlowski2001,Holland2002}. Similar timescales may be needed throughout the Universe, depending on biochemistry. \\
\indent However, the environment on small, icy exomoons may be subject to change on timescales of millions rather than billions of years. Tidal heating may at some given time maintain a liquid ocean, but if that heat disappears the moon is frozen solid. \cite{DrisBarnes2015} studied the tidal evolution of planetary orbits in M dwarf systems (similar in scale to the Jovian system) and find that the closer in a planet orbits, independent of initial eccentricity, the sooner their orbits circularize. For a planet with a 0.01 AU semi-major axis (approximately the semi-major axis of Titan around Saturn), an orbit with an initial eccentricity of 0.5 circularizes entirely within 1 million years after intense tidal heat dissipation. Once the orbit is circularized tidal heating drops to zero; in the absence of major exogenic heating contributions the moon is then frozen solid. Mean motion resonances are believed to sustain tidal forces on Io, Europa, and Enceladus \citep{Yoder1979,Porco2006}, but these pumping mechanisms do not always endure: the moons of Uranus are thought to have been in multiple mean motion resonances but are no longer \citep{TitteWisdom1990}. \\
\indent Given the narrow band of molten states for Enceladus and Rhea, even minor orbital evolution may already see the moon freeze solid. Larger, more distant moons such as Titania and Oberon are massive enough to maintain liquidity even at very small eccentricities (see the eccentricity plots in Figures \ref{fig:singleTit} and \ref{fig:singleOb}), but we estimate their radiogenic background heating and NH$_3$ (or similar pollutant) fractions to be higher and their regolith blankets to be thicker. Given the possibility that Enceladus and the other icy Saturnian moons formed as little as 100 Myr ago \citep{Cuk2016} we must acknowledge that while such environments could be habitable now, they need not be or have been for long. It may then be prudent to distinguish between transient habitable worlds, which through tidal effects are only habitable for a geologically brief interlude, and permanently habitable worlds such as Earth, which maintain that status on billion year timescales.

\subsection{Implications for exomoons}
\label{ssec:disc/impforexoms}
Our fiducial moons are all Solar System moons, since at the time of this writing we have no alternatives. However, our aim remains to determine the habitability of exomoons under different environmental circumstances. We believe that our approach is general enough that it may be applied successfully to any set of fiducial values; our findings are snapshots of the (at least) seventeen-dimensional parameter space describing subsurface habitability within which any moon can be placed. We could continue sampling this space at arbitrary points to obtain a more complete picture; however, Equation \ref{eq:meltdepthan} gives us an impression of what needs to be known about how initial moon observables relate to the melting depth and can be applied to both Solar System moons and exomoons alike. We therefore regard Equation \ref{eq:meltdepthan} as the main result of this work. If, in the future, the properties of some exomoon become known, we can use this equation to gain an indication of whether that particular exomoon could possibly sustain a subsurface ocean.

\subsection{Model limitations}
\label{ssec:disc/modlims}
This work does not study the long-term evolution of exomoon systems. Our model provides a snapshot of an icy moon but does not predict its state in several million years, or several million years ago. Its orbital characteristics may evolve (with profound effects on tidal heating, particularly if the maximum melting criterion is trespassed) or it may be destroyed entirely. This limits our ability to draw definitive conclusions regarding the habitability of such environments on anything but transient timescales. Such developments should be studied in further research. \\
\indent Secondly, our phenomenological approach to tidal heating is a simplified model which has in accuracy been superseded by viscoelastic models (see Section \ref{ssec:physb/aptoth}; also \citealt{Moore2003}, \citealt{Henning2009}, \citealt{DobTurn2015}). While specifically the (still simplified) Maxwell model has found much literature use, more sophisticated models exist (such as Burger's and Andrade; see \citealt{RenaudHenning2018}). Key is that viscoelastic models allow for molten interiors; not implementing this limits our model to moons small enough to maintain a uniform, nonmolten interior (less than $10^{22}$ kg, i.e., about half the mass of Triton). Since small, icy moons such as Enceladus possess uniform, nonmolten interiors, viscoelastic tides should not strongly affect our predictions for heat dissipation in this regime. \\
\indent Thirdly, we are limited by our knowledge of several key ingredients of our melting depth model. For more accurate predictions we must first properly constrain the tidal efficiency factor $\Phi$ of our fiducial moons and determine how it scales with readily observable characteristics. The regolith cover of icy moons also needs to be investigated, so we may put better limits on its thickness. \\
\indent Fourthly, we do not take into account the effects of asymmetric structure and, though this applies only to geologically active exomoons, cryovolcanism. Both phenomena occur on Enceladus, with the icy envelope presumably shifted relative to its core \citep{Nimmo2018} and active geysers on its south pole. These two phenomena may also influence each other: we speculate that an asymmetric shell on a tidally active exomoon may result in a crust so thin in certain regions that cryovolcanic activity there follows. It was pointed out to the authors by T. Steinke and M. Navarro from TU Delft that up to 90\% of Io's internal heat may be dissipated via advective (volcanic) processes; while Enceladus has far less internal heating than Io, the amount of heat lost via advection may thus still be appreciable and should be considered in future approaches (see also \citealt{Hussmann2002}). \\
\indent Fifthly, this work does not consider tidal heat dissipation in the moon's ocean or ice crust. We assume all tidal heat to be dissipated in the (rocky and porous) core and mantle; this is not necessarily the case. Appreciable amounts of tidal heat can still be dissipated in the ocean; \cite{Tyler2008} states that Enceladus' entire excess heat flux may be explained by tidal dissipation in its ocean alone rather than its core. For moons with a thick ice crust, such as Oberon and Titania, tidal heat dissipated in the crust may also play a role \citep{WilKers2018}.

\subsection{Observability}
\label{ssec:disc/obs}
While our model provides an outlook into the parameter space of potentially habitable, icy exomoons, its implications are not yet testable. While there is at least one candidate (Kepler 1625b-I; see \citealt{TeachKip2018}, though doubts have been expressed on this result; see, e.g., \citealt{Rodenbeck2018}), no exomoon detection has yet been confirmed, let alone an exo-ocean detection. The challenge of exomoon detection (in particular for small, icy exomoons such as Enceladus) lies in both their size and orbit. The smallest exoplanets currently confirmed have radii of order $\sim$0.1 Earth radii and are on very tight orbits (e.g., \citealt{Vanderburg2015}); while Rhea's radius is of this order, Enceladus' radius is three times smaller than this ($\sim$250 km), and icy bodies can not maintain their volatiles interior to the snow line. Additionally, regular transit photometry, such as for exoplanets, can not easily be performed for exomoons since the moon's phase relative to the planet is different during each transit, thereby complicating lightcurve folding. Transit photometry around directly imaged exoplanets may be done in the future but is currently still beyond our abilities. When this becomes possible, transit spectra of possible lunar atmospheres or plumes can be used to infer the presence of subsurface oceans. \\
\indent The upcoming PLATO mission\footnote{\url{https://platomission.com/}} is designed to observe planets beyond the snow lines of M and K dwarf systems, which is the type of host planet that we are interested in. In addition, PLATO may be able to detect large exomoons through transit timing or duration variations, whereby the gravitational pull of the companion perturbs the timing or length of the exoplanet's transit \citep{PLATO2014}. While they stress the difficulty of detecting exomoons around widely orbiting hosts, \cite{PLATO2014} also indicate that exomoon radius and atmospheres could be studied during exoplanet-exomoon pair transits. However, while in this fashion Galilean-size exomoons may be found, it is unlikely that Enceladus-size moons could be similarly detected. Enceladus is $\sim$1000 times less massive than Ganymede; Enceladus' projected surface area onto its host star would be $\sim$100 times smaller than Ganymede's. Rhea, Titania and Oberon are $\sim$100 times less massive than Ganymede; their projected surface areas would be $\sim$10 times smaller than Ganymede's. We would need at least a tenfold increase in telescopic sensitivity to detect Rhea analogs and a hundredfold increase to observe an exo-Enceladus. For transit timing or duration variations, the prospect is even more demanding, with a required thousandfold increase in sensitivity. Once a small exomoon is found, we could then use our model to investigate whether an ocean would be possible on the inferred moon, but we would then need to observe the moon's plumes to provide confirmation. \\
\indent In short, the detection of exomoons of the scale studied in this paper is still remote; however, the detection of exomoons in general is plausible in the near future, once PLATO becomes operational \citep{PLATO2014}. For example, the lower limit on transit depth for present-day exoplanet detections is of order $\sim$20 ppm (see, e.g., \citealt{Barclay2013}, figure 1a); a Rhea-sized body transiting before a Sun-size star would have a transit depth of $R_{Rhea}^2/R_{\odot}^2\sim$1 ppm, whereas for a Ganymede analog the transit depth would be $\sim$15 ppm. By comparison, PLATO's minimal transit depth goes down to $\sim$10 ppm for 9$^{th}$ magnitude stars and brighter \citep{Catala2010}, so the Ganymede analog could be detected.

\section{Conclusions}
\label{sec:conclusions}
The habitability of exomoons as we defined it depends on whether liquid water can exist below the surface. Subsurface oceans analogous to what is believed to exist on Enceladus (and other Solar System moons) may thus provide a habitat for extraterrestrial life. We have found that beyond the snow line, these oceans may exist largely irrespective of distance from the host star, rendering the exomoon subsurface habitable zone arbitrarily large. Since (in our Solar System) there are more moons than planets, habitable planets may be vastly outnumbered by habitable moons. Moons have several heat sources which planets lack, making them less dependent on stellar illumination: tides from their host planet, assuming the moon's eccentricity is nonzero, may dissipate ample heat in a moon's interior to sustain a subsurface ocean. \\
\indent We have studied under which conditions subsurface oceans can exist. To this end, assuming conductive heat transfer through the moon's ice shell only, we derived an analytic expression for the melting depth, Equation \ref{eq:meltdepthan}, which depends on seventeen parameters. Prime among these are the moon's mass, semi-major axis, orbital eccentricity, regolith cover, NH$_3$ fraction, and tidal efficiency factor; the planetary mass and semi-major axis; and the stellar luminosity. This means that given some set of parameters for some hypothetical moon, we can (optimistically) estimate at what depth liquid water may occur. \\
\indent Using Enceladus and Rhea by Saturn, and Titania and Oberon by Uranus as fiducial cases, we have explored the range of model parameters. Figures \ref{fig:medeEnc} through \ref{fig:medeDeadRhea} show that subsurface oceans barely rely on stellar illumination: our fiducial moons' melting depths vary little with distance from the Sun. More important is a source of internal heat that serves to keep the ocean liquid. This source may be tidally induced or radiogenic. \\
\indent Given a sufficiently thick regolith blanket and a high tidal efficiency factor, it is possible to maintain a subsurface ocean at arbitrary distances from the host star. With their dark surfaces, Titania and Oberon radiate little heat and may retain their internal energy on long timescales; Enceladus and Rhea both receive tidal heat from Saturn. Despite this, Rhea seems frozen, yet if we change the (poorly constrained) tidal efficiency factor appropriately it may sustain a subsurface ocean as well, or even be fully desiccated; see Figure \ref{fig:medeDeadRhea}. An NH$_3$ mass fraction of several percents also contributes by lowering the melting point of the ice crust but may end up rendering the ocean saturated with NH$_3$ and thus toxic to life as we know it. We must note that our models provide only a snapshot of exomoons at a certain point in time; in no way can we claim that exomoons may retain such habitable environments forever. \\
\indent This leads us to conclude that habitable exomoons (surface and subsurface) may be found scattered throughout exoplanetary systems. The circumstellar subsurface habitable zone for moons extends arbitrarily farther out than for planets. Hence, the primary limit on icy moon habitability is their growth from the preceding planetesimal disk. We should not find subsurface habitable moons closer to the star than the snow line; stellar illumination would render them desiccated through sublimation of the ice crust once they trespass this boundary. Hence, the snow line can be regarded as a circumstellar subsurface habitable edge. Like \cite{HellerBarnes2013}, we also find a circumplanetary habitable edge, although we believe (at least when it concerns subsurface habitability) it better regarded as a maximum melting criterion because it depends on multiple parameters. Regardless, we speculate that subsurface habitable exomoons may populate the parameter space from the least massive rounded moons (Mimas and Enceladus) to at least Titania, and possibly well beyond. \\
\indent Future exomoon hunters may use Equation \ref{eq:meltdepthan} to obtain an estimate of whether their newly discovered worlds are capable of sustaining a subsurface ocean. Since we assume conduction only, such an estimate would form a lower bound; a more realistic melting depth would, if volcanism is taken into account, always end up deeper. However, if volcanism is at all present this already suggests the presence of a subsurface ocean or lakes (to supply the cryovolcanic ejecta) and our model would not be required to infer an ocean's presence to begin with. In the end, our model is not to be taken as a detailed physical model of particular (exo)moons but as a general purpose model useful for putting preliminary boundaries on what is possible, applicable to any small, icy exomoon. It allows users to gain a swift, qualitative impression of what parameters are most important to a small, icy exomoon's habitability. A specifically tailored model can then be subsequently used to obtain a quantitative understanding of the particular moon. \\
\indent Our study can be improved upon by better study of ice thermal conductivity behavior under vacuum circumstances; by better constraints on regolith thickness, composition, and crustal NH$_3$ mass fraction; by improved determination of the tidal efficiency factor $\Phi$ and its coupling to internal structure, orbital characteristics, and observable qualities of exomoons; and by physical exploration of our fiducial moons so that we may compare our predictions with empirical data. Similar studies could also be carried out into the habitability of moonmoons, which we expect may have more complex tidal behavior. The authors would like to express their hope that this research, by demarcating areas of interest, in some way contributes in the hunt for exo-oceans.

\section{Acknowledgements}
\label{sec:acknowledgements}
The authors would like to thank T. Steinke and M. Rovira-Navarro of TU Delft for the useful discussion. The authors would also like to thank R. Heller of the Max Planck Institute for Solar System Research and R. Barnes of the Virtual Planetary Laboratory for their useful comments.

\bibliographystyle{aa}
\bibliography{aanda.bib}

\appendix
\section{Derivation of melting depth}
\label{app:derivmeltdep}
Recall from Section \ref{ssec:physb/heattrans} that the surface temperature is given by Equation \ref{eq:surftemp}. If this surface temperature exceeds 150 K (the snow line; see \citealt{Armitage2016}), we assume the water ice on the surface to sublimate, leaving the moon desiccated; if not, we compute the temperature underneath the insulating regolith layer $D_{rego}$. Since, assuming symmetric heat loss, the endogenic flux at some depth $r$ must be

\begin{align*}
    F_{endo}(r) = \dfrac{\dot{E}_{endo}}{4\pi r},
\end{align*}

wherein $\dot{E}_{endo}$ is the total generated endogenic heat, we can use equation \ref{eq:tempgrad} to obtain:

\begin{align*}
    T_{rego} &= T_{surf} + \Delta T_{rego} \\
    &= T_{surf} + \dfrac{D_{rego}\cdot\dot{E}_{endo}}{4\pi \left(R_s - D_{rego}\right)^2 k_{rego}},
\end{align*}

wherein $k_{rego}$ is the thermal conductivity of the porous regolith layer; since this layer consists of finely grained material (shattered ice, snow, dust), the majority of its volume is vacuum (in the absence of an atmosphere) and heat transport is therefore dominated by radiative processes. These are very inefficient, so the value of $k_{rego}$ is taken to be 0.001 W m$^{-1}$ K$^{-1}$ (based on \citealt{YuFa2016}). \\
\indent Subsequently, we compute the temperature at some depth $D$ into the solid ice shell. Recall that the thermal conductivity of ice is given by Equation \ref{eq:icecond}, as found by \cite{AndersInaba2005}. Let us then subdivide the ice shell in layers of thickness $\Delta D$, each with a higher temperature than the one above and thus a different thermal conductivity. Suppose the temperature in layer $n$ is $T_n$; following Equation \ref{eq:tempgrad}, the temperature in layer $n+1$ at depth $D_{n+1}$ must then be

\begin{align*}
    T(D_{n+1}) &= T(D_n) + \Delta T(D_{n+1}) \\
    &= T(D_n) + \dfrac{\Delta D\cdot\dot{E}_{endo}}{4\pi \left(R_s - D_{n+1}\right)^2 k(T(D_{n+1}))},
\end{align*}

wherein $D_n$ is the depth of layer $n$. We divide $\Dot{E}_{endo}$ by the surface area of each layer separately since, as the surface area shrinks, the amount of heat escaping per surface area must go up. Doing this iteratively, we obtain (by setting $T_{rego}=T_0$):

\begin{align*}
    T(D_N) = T_{rego} + \sum^{N}_{n=1}\dfrac{\Delta D\cdot\dot{E}_{endo}}{4\pi \left(R_s - D_n\right)^2 k(T(D_n))}.
\end{align*}

If $\Delta D$ tends to zero, this becomes the following integral:

\begin{align*}
    T(D) = T_{rego} + \int^{D}_{D_{rego}}\dfrac{\dot{E}_{endo}}{4\pi \left(R_s - D'\right)^2 k(T(D'))}dD'.
\end{align*}

Where the superscript $'$ is added to distinguish between the actual target depth $D$ and the integration variable $D'$. This integral can not be immediately solved since $k$ is dependent on $T$, which is dependent on $D$ in an unknown way. Hence, on both sides, we take the derivative with respect to $D$:

\begin{align*}
    \dfrac{dT}{dD} &= \dfrac{\dot{E}_{endo}}{4\pi \left(R_s - D\right)^2 k(T)} \\
    \xrightarrow{} k(T)dT &= \dfrac{\dot{E}_{endo}}{4\pi} \left(R_s - D\right)^{-2} dD.
\end{align*}

Let us define (for notational ease) the constant factor $\dot{E}_{endo}/4\pi = X$. Next, to find the melting depth, we integrate on both sides from $T_{rego}, D_{rego}$ to $T_l, D_l$ ($l$ for liquid):

\begin{align*}
    \int^{T_l}_{T_{rego}}k(T)dT = X \int^{D_l}_{D_{rego}}\left(R_s - D\right)^{-2} dD.
\end{align*}

Filling in Equation \ref{eq:icecond} on the left, we find that

\begin{align*}
    \int^{T_l}_{T_{rego}}k(T)dT &= \left[632\ln{T} + 0.38T - 0.00985T^2\right]^{T_l}_{T_{rego}} \\
    &= K(T).
\end{align*}

Let us define the function $632\ln{T} + 0.38T - 0.00985T^2 = K(T)$, the integrated conductivity. Then integrating the right side gives

\begin{align*}
    X \int^{D_l}_{D_{rego}}\left(R_s - D\right)^{-2} dD = X\left[\left(R_s - D\right)^{-1}\right]^{D_l}_{D_{rego}}.
\end{align*}

So we find that

\begin{align*}
    K(T_l) - K(T_{rego}) = X\left(R_s - D_l\right)^{-1} - X\left(R_s - D_{rego}\right)^{-1},
\end{align*}

which we can solve for $D_l$, thus (after resubstituting $X$) finally yielding

\begin{align*}
    D_l = R_s - \left[4\pi\dot{E}_{endo}^{-1}\left(K(T_l) - K(T_{rego})\right) + \left(R_s - D_{rego}\right)^{-1}\right]^{-1},
\end{align*}

which is our analytic expression for the melting depth.

\newpage\onecolumn
\section{Full melting depth grids}
\label{app:fullmdgrids}

\begin{figure}[h]
    \includegraphics[width=\columnwidth]{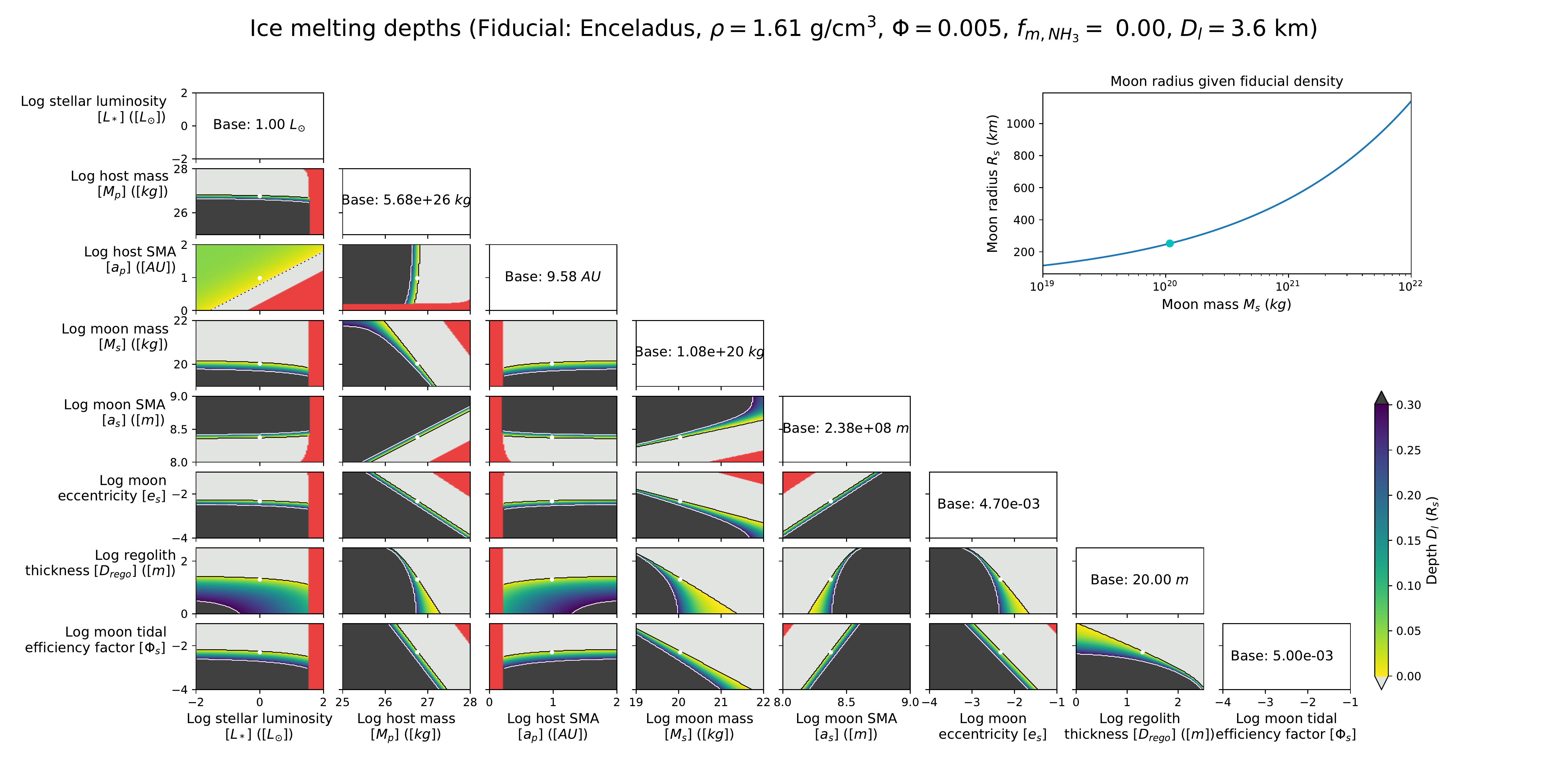}
    \caption{Full melting depth parameter grid for Enceladus. \label{fig:medeEnc}}
\end{figure}

\begin{figure}[h]
    \centering
    \includegraphics[width=\columnwidth]{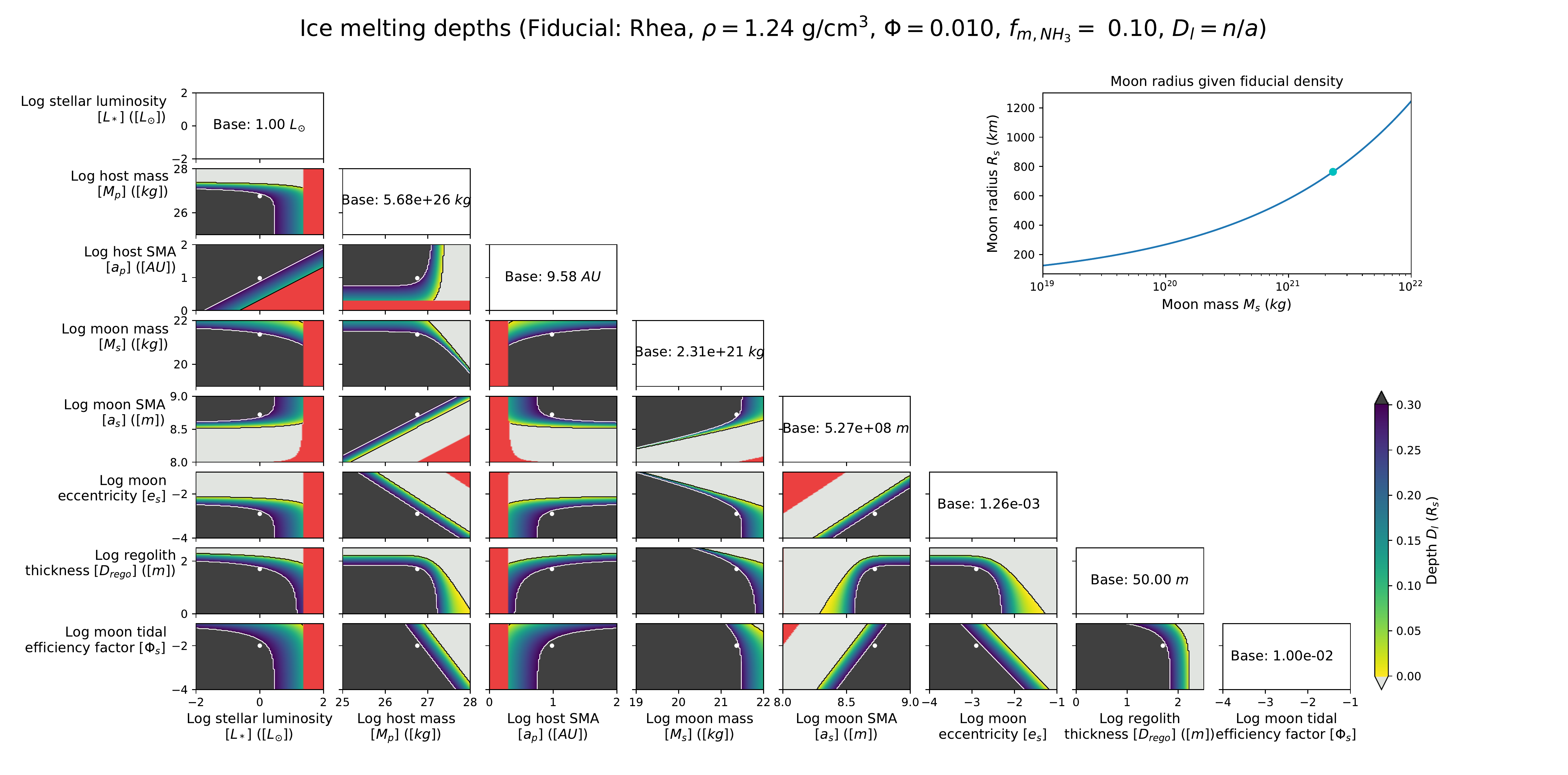}
    \caption{Same as previous figure, but for Rhea. \label{fig:medeRhea}}
\end{figure}

\begin{figure}[h]
    \centering
    \includegraphics[width=\columnwidth]{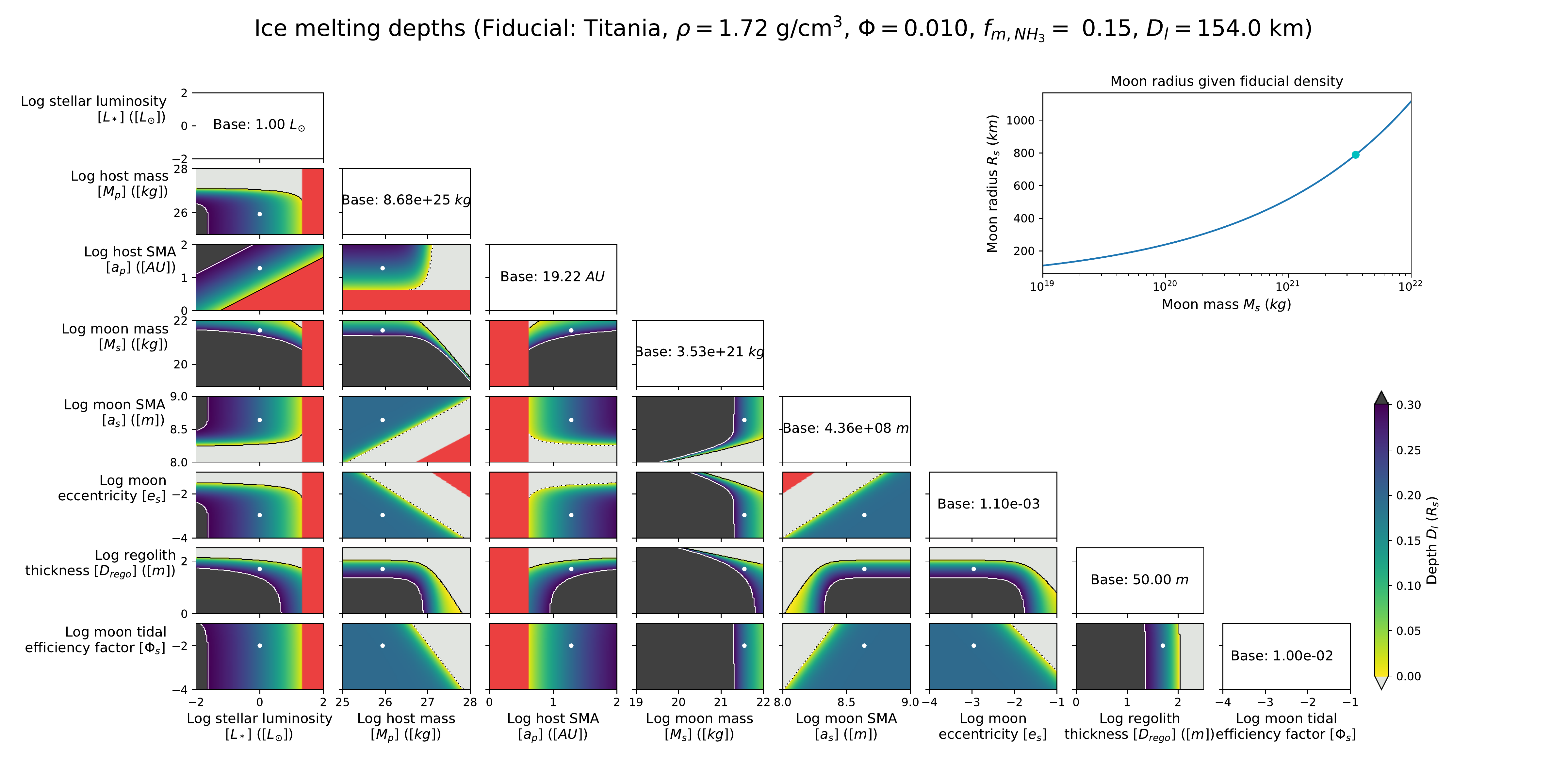}
    \caption{Same as previous figure, but for Titania. \label{fig:medeTit}}
\end{figure}

\begin{figure}[h]
    \centering
    \includegraphics[width=\columnwidth]{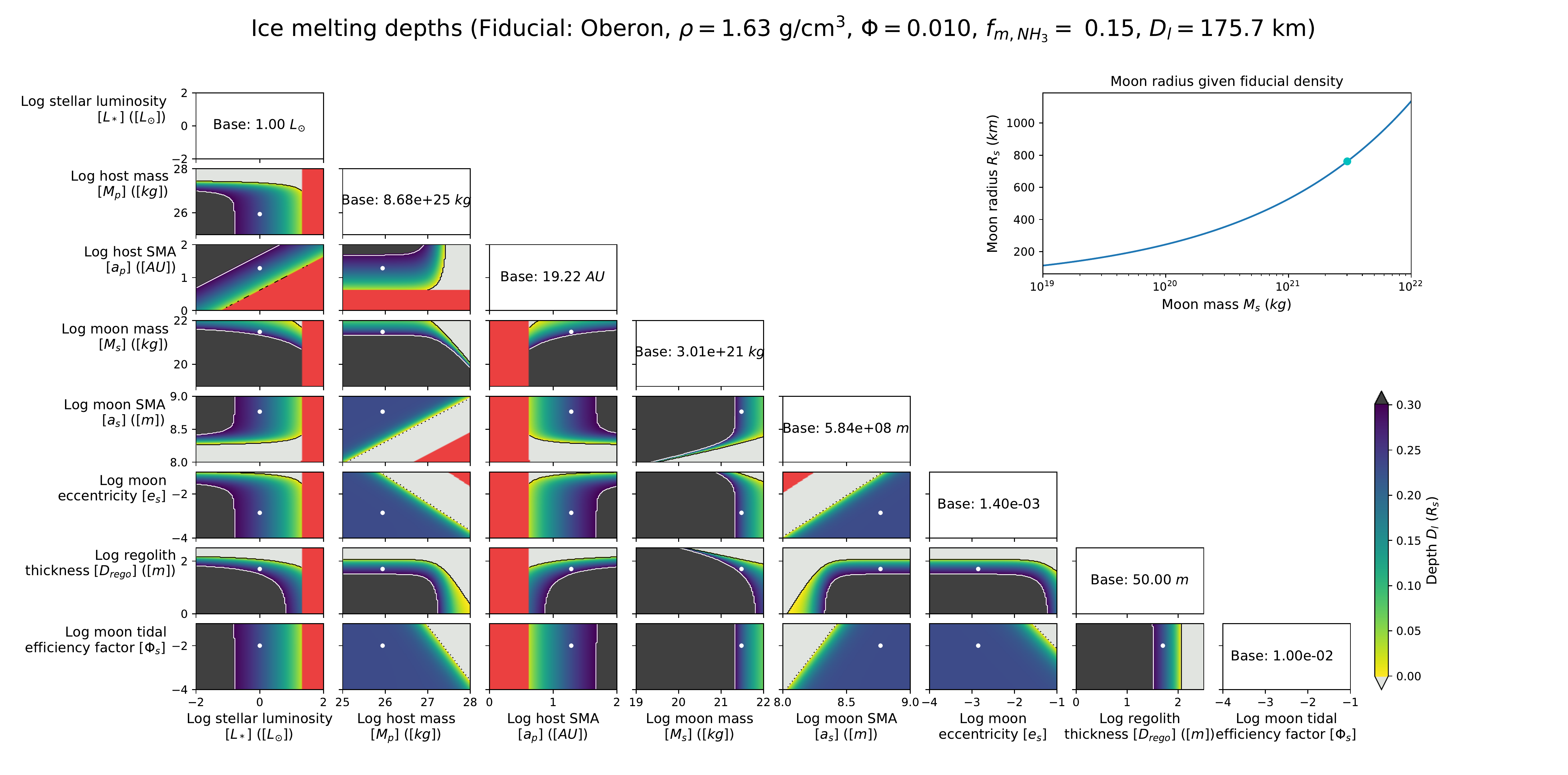}
    \caption{Same as previous figure, but for Oberon. \label{fig:medeOber}}
\end{figure}

\begin{figure}[h]
    \centering
    \includegraphics[width=\columnwidth]{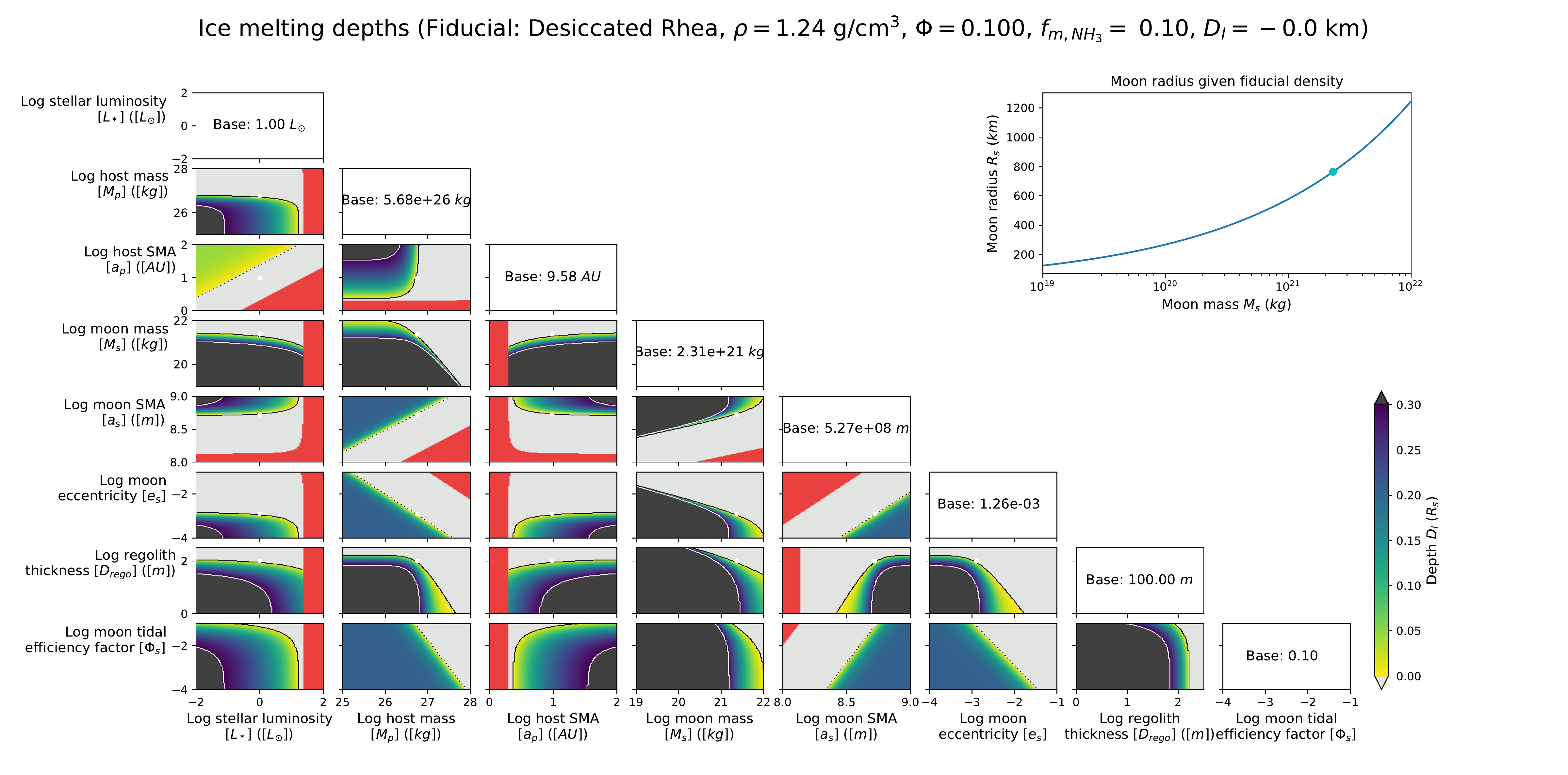}
    \caption{Same as previous figure, but for a hypothetical desiccated Rhea. A tenfold increase in the tidal efficiency factor and a doubling of the regolith layer, while extreme and probably unphysical, can transform even frozen Rhea into a desiccated rock. \label{fig:medeDeadRhea}}
\end{figure}

\begin{figure}[h]
    \centering
    \includegraphics[width=\columnwidth]{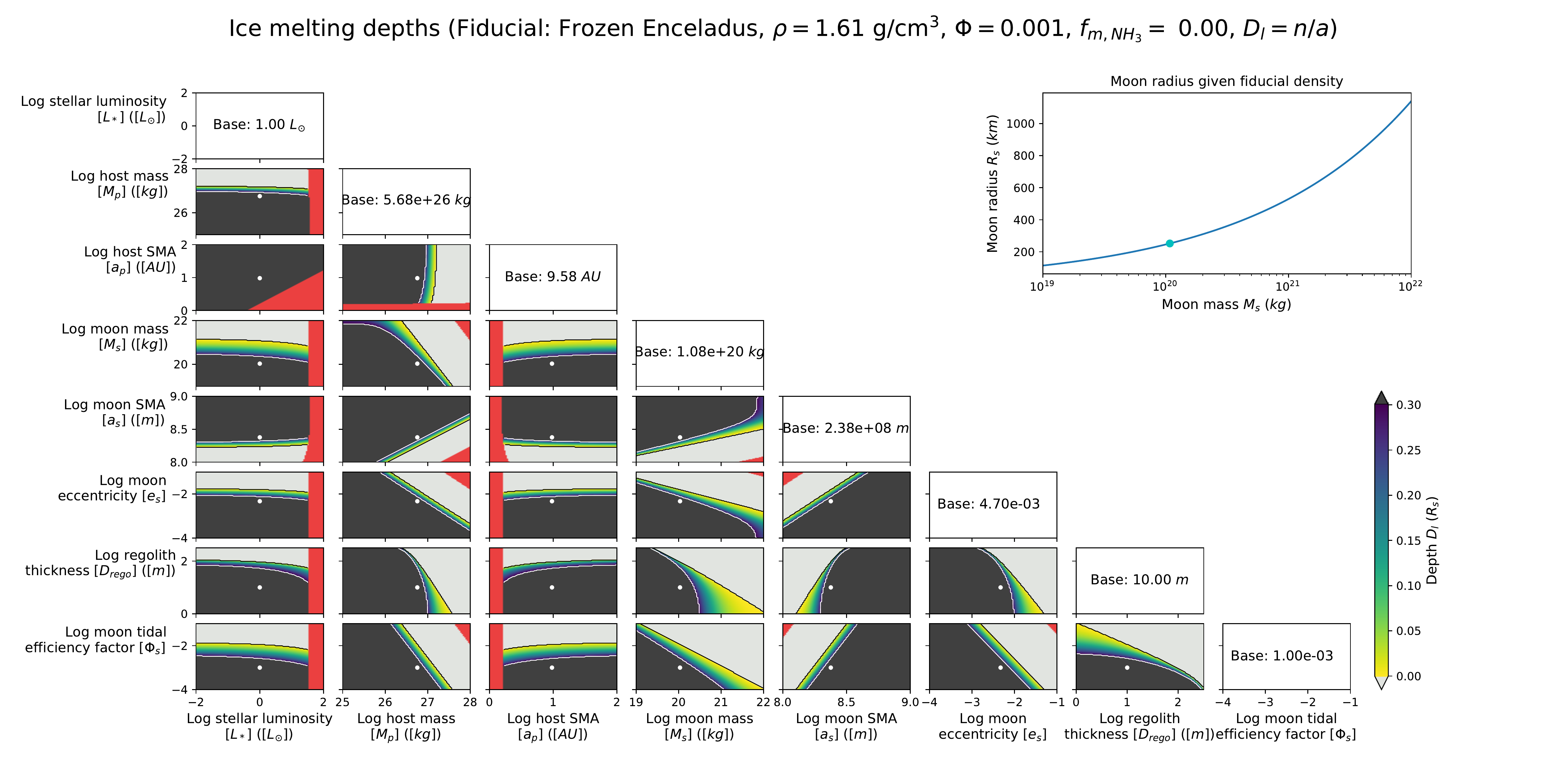}
    \caption{Same as previous figure, but for a hypothetical frozen Enceladus. Dividing the tidal efficiency factor by 5 and halving the regolith blanket results in Enceladus cooling dramatically, turning it into solid ice. \label{fig:medeFrozenEnc}}
\end{figure}

\end{document}